\documentclass[useAMS,usenatbib,usegraphicx,usedcolumn]{mn2e}
\usepackage{lscape}
\usepackage{graphicx}
\usepackage{txfonts}
\usepackage{natbib}
\usepackage{lscape}
\input{psfig.sty}

\newcommand{\pedix}[2]{\ensuremath{#1_{\,\mbox{\scriptsize #2}}}}
\newcommand{\apix}[2]{\ensuremath{#1^{\,\mbox{\scriptsize #2}}}}
\newcommand{\pedap}[3]{\ensuremath{#1_{\,\mbox{\scriptsize #2}}^{\,\mbox{\scriptsize #3}}}}
\newcommand{\pedSM}[2]{\ensuremath{#1_{\,\mbox{\tiny #2}}}}
\newcommand{\errUD}[2]{\ensuremath{^{+#1}_{-#2}}}
\newcommand{\ie}{{i.e.}}
\newcommand{\eg}{{e.g.}}
\newcommand{\mydag}{\ensuremath{\mbox{\dag}}}
\newcommand{\ion}[2]{\ensuremath{\mbox{#1~{\sc #2}}}}
\newcommand{\feka}{\ensuremath{\mbox{Fe~K}\alpha}}
\newcommand{\fekb}{\ensuremath{\mbox{Fe~K}\beta}}
\newcommand{\oviii}{\ensuremath{\mbox{\ion{O}{viii}}}}
\newcommand{\neix}{\ensuremath{\mbox{\ion{Ne}{ix}}}}
\newcommand{\Lya}{\ensuremath{\mbox{Ly}\alpha}}    
\newcommand{\Ly}[1]{\ensuremath{\mbox{Ly}#1}} 
\newcommand{\asca}{{\emph{ASCA}}}
\newcommand{\cgro}{\emph{Compton Gamma Ray Observatory}}
\newcommand{\chandra}{{\emph{Chandra}}}
\newcommand{\comptel}{{\emph{COMPTEL}}}
\newcommand{\egret}{{\emph{Egret}}}
\newcommand{\fermi}{\emph{Fermi}}
\newcommand{\integral}{\emph{INTEGRAL}}
\newcommand{\osse}{\emph{OSSE}}
\newcommand{\rxte}{{\emph{RXTE}}}
\newcommand{\sax}{\emph{Beppo}SAX}
\newcommand{\suzaku}{\emph{Suzaku}}
\newcommand{\swift}{{\emph{Swift}}}
\newcommand{\xmm}{{XMM-\emph{Newton}}}
\renewcommand{\sun}{\ensuremath{\odot}}
\newcommand{\lum}{\ensuremath{\mbox{ergs~s}^{-1}}}
\newcommand{\flux}{\ensuremath{\mbox{ergs~cm}^{-2}\mbox{~s}^{-1}}}
\newcommand{\nh}{\ensuremath{\mbox{cm}^{-2}}}
\newcommand{\nhsym}{\ensuremath{N_{\mbox{\scriptsize H}}}}
\newcommand{\arcdeg}{\ensuremath{^{\circ}}}
\newcommand{\kev}{\ensuremath{\,\mbox{\scriptsize keV}}}
\newcommand{\normGAUSS}{\ensuremath{\mbox{photons~cm}^{-2}\mbox{~s}^{-1}}}
\newcommand{\chidof}{\ensuremath{\chi^2/\mbox{d.o.f.}}}
\newcommand{\dchidof}{\ensuremath{\Delta\chi^2/\Delta\mbox{d.o.f.}}}
\newcommand{\treca}{3C~111}
\newcommand{\trecb}{3C~382}

\title[The High Energy view of the Broad Line Radio Galaxy 3C~111]
{The High Energy view of the Broad Line Radio Galaxy 3C~111}
\author[L. Ballo et al.]
{L.~Ballo$^{1}$\thanks{E-mail:
ballo@ifca.unican.es (LB)}, V.~Braito$^{2}$, J.~N.~Reeves$^{3}$, 
R.~M.~Sambruna$^{4}$, and F.~Tombesi$^{5,6}$
\\
$^{1}$ Instituto de F\'{\i}sica de Cantabria (CSIC-UC), Avda. Los Castros s/n 
(Edif. Juan Jord\'a), E-39005 Santander, Spain \\
$^{2}$ Department of Physics and Astronomy, University of Leicester, University Road,
Leicester LE1 7RH, UK \\
$^{3}$ Astrophysics Group, School of Physical and Geographical Sciences, Keele University,
Keele, Staffordshire ST5 5BG, UK \\
$^{4}$ Department of Physics and Astronomy, MS 3F3, 4400 University Drive, George Mason
University, Fairfax, VA 22030, USA \\
$^{5}$ X-ray Astrophysics Laboratory and CRESST, NASA/Goddard Space Flight Center,
Greenbelt, MD 20771, USA \\
$^{6}$ Department of Astronomy, University of Maryland, College Park, MD 20742, USA
}
\begin{document}

\date{Accepted 2011 August 12.  Received 2011 August 11; in original form 2011 May 24}

\pagerange{\pageref{firstpage}--\pageref{lastpage}} \pubyear{XXXX}

\maketitle

\label{firstpage}

\begin{abstract}
We present the analysis of \suzaku\ and \xmm\ observations of the broad-line
radio galaxy (BLRG) \treca.
Its high energy emission shows 
variability, a harder continuum with respect to the radio quiet AGN population,
and weak reflection features.
\suzaku\ found the source in a minimum flux level; a comparison with the \xmm\ data implies an increase of a
factor of $2.5$ in the $0.5-10\,$keV flux, in the $6\,$months separating the two observations.
The iron K complex is
detected in both datasets, with rather low equivalent width(s).
The intensity of the iron K complex does not respond to the change in continuum 
flux.
An ultra-fast, high-ionization outflowing gas is clearly detected in the XIS data; the absorber is most likely unstable.
Indeed, during the \xmm\ observation, which was $6\,$months after, the absorber was not detected.
No clear roll-over in the hard X-ray emission is detected, probably due to the emergence of the
jet as a dominant component in the hard X-ray band, as suggested by the detection above 
$\sim 100\,$keV with the GSO on-board \suzaku, 
although the present data do not allow us to firmly constrain the relative
contribution of the different components.
The fluxes observed by the $\gamma$-ray satellites {\it CGRO} and \fermi\ would be compatible with the 
putative
jet component if peaking at energies $E\sim 100\,$MeV.
In the X-ray band, the jet contribution to the continuum starts to be significant only above $10\,$keV.
If the detection of the jet component in \treca\ is
confirmed, then its relative importance in the X-ray
energy band could explain the different observed properties in
the high-energy emission of BLRGs, which are otherwise similar in their
other multiwavelength properties.
Comparison between X-ray and $\gamma$-ray data taken at different epochs suggests that the strong variability observed 
for \treca\ is probably driven by a change in the primary continuum.
\end{abstract}



\section{Introduction}\label{sect:intro}

The unified model of Active Galactic Nuclei (AGN) predicts that
the ability of an accretion disk-black hole system to produce powerful relativistic jets 
is the main reason of the observed differences between radio loud (RL) and radio quiet (RQ) sources.
Geometrical effects, through the relative inclination of an axisymmetric 
system with respect to the line of sight, can explain the majority of the differences 
between the various subclasses \citep{antonucci93,urry95}.
Although this model does work well to first order, it leaves still open a number of problems.
 
In particular, a key question still to be fully answered is why
powerful relativistic jets are produced
only in $10-20$\% of AGN,
and if and how this is related to the structure of
the accretion flow in the RL and RQ sources.
In other words, an often debated issue is how accretion and ejecta in AGN are linked,
and how these mechanisms work under different physical conditions
(\eg, \citealt{bz,bp,sikora07,garofalo09,garofalo10,tchekhovskoy10}; for a parallel with X-ray binaries, see
\citealt{fender10}).
A closely related question concerns the evolution of the jets, and its link with 
the observational subclasses of RL.
It is commonly accepted that the relativistic jet originates in 
the innermost regions,  and as the angle
between the jet axis and the line of sight decreases,  the importance of  
its emission increases, due to beaming effects.
This simple picture does not account for the whole RL phenomenology, different 
global properties being associated to sources with different radio morphology
\citep[\eg,][]{hardcastle07,daly09a,daly09b}.

A key observational challenge in RL AGN is to disentangle the jet contribution from the
disk emission, which could allow us to better understand what are the main 
differences in the nuclear regions between
RL and RQ sources.  Previous X-ray observations of  RQ and RL 
objects  established that these sources exhibit spectra with 
subtle but significant differences \citep[\eg,][]{zdziarski01,sambruna02}.
The X-ray emission of RQ objects appears to be described at first order   
by a primary power-law continuum, with
features resulting from the reprocessing of this  primary continuum from cold 
and warm gas, like the iron K line complex
at $6\,$keV, the reflection component, and often ionized absorption and emission 
features.
The main observed difference between RL and RQ sources is that 
the features due to reprocessing in RL AGN  appear to be weaker compared to 
the RQ population \citep[and references therein]{ballantyne07}.

To account for these properties, several scenarios have been suggested.
Highly ionized accretion disks, different from the standard, cold disk typical
of Seyferts, could be obtained if the accretion rates are high
\citep{ballantyne02}.
A difference in the inner accretion disks of BLRGs and RQs could result in a
small solid angle subtended in RLs by the disk to the primary X-ray source
\citep{eracleous00}.
On the other hand, assuming the same disk structure in both populations, 
dilution by non-thermal jet emission could weaken the reprocessing features
\citep{grandi02}.

It is however important to notice that  in the last few years \xmm\ and \suzaku\
are changing this  simplistic view: observations of large samples of 
RQ AGN are showing a large spread in the X-ray properties of this class, with 
a wider range of   continuum slopes and with several cases of   sources having
little or no reflection.
It is thus emerging that  the distinction between the two classes is not so
sharp, but rather that RL objects
populate one end of the distribution for Seyferts and quasars \citep[QSOs;][and references therein]{sambruna09}.

Strong signatures of low- and high-velocity outflows are rather common in RQ
\citep[e.g., ][]{turner09}.
Warm absorbers (outflowing with velocities $v\sim 100-1000\,$km/s) are detected in the soft X-ray spectra of more
than half of local Seyfert galaxies \citep[\eg,][]{crenshaw03,blustin05}.
Their location at $\sim 1-100\,$pc suggests a possible association with the optical Broad Line Region (BLR) or Narrow
Line Region.
Ultra-fast outflows, showing velocities $v > 10000\,$km/s, have been found to be present in $\sim 40$\% of local 
Seyferts \citep{tombesi10sey}. 
Their location on sub-pc scales suggests a direct association with accretion disk winds/outflows.

Regarding RL objects, only in the last few years have sensitive and broad-band observations started to detect disk 
winds/outflows in Broad Line Radio Galaxies (BLRGs).
\chandra\ and \xmm\ observations of \trecb\ revealed the presence of a warm absorber with outflow-like 
properties \citep{reeves09,torresi10}, while
3C~445 shows signatures of soft X-ray photoionized gas in emission and
absorption \citep[see][]{sambruna07,reeves10,braito11}.
Our \suzaku\ observations provided evidence for ultra-fast disk outflows in 3/5 BLRGs 
\citep[blueshift velocity $\pedix{v}{out}\sim 0.04-0.15\,c$, and mass outflow
rates comparable to the accretion rates;][]{tombesi10}.

Differences in terms of the strength of the reflection features, as well as the
presence or lack of warm absorbers or high-velocity outflows, in sources showing different radio properties
is one of the natural consequences of the so-called ``gap-paradigm'' \citep{garofalo09,garofalo10}.
Its main ingredient is the relative orientation of the disk and the black hole spin; 
the ability of the source of producing powerful jets is related to retrograde systems.
In particular, this model accounts for the weaker reflection features observed 
in BLRGs, a natural consequence of the larger gap between the innermost stable 
circular orbit of the disk and the black hole in retrograde sources 
than in prograde objects.
At the same time, the model provides a simple interpretation for the presence or
absence of signatures of disk winds, depending on the size of the gap
region and the accretion efficiency.

An additional complication in the study of the properties of the accretion flow in RL and RQ, is 
the presence in the former of emission due to the jet, that can mask the thermal emission (directly observed and/or
reprocessed) from the disk.
From the observational point of view, only  through a  wide energy coverage and  
good  sensitivity
at medium-hard X-rays, can we attempt to disentangle the contribution of the 
jet and the disk emission.
The request of both simultaneous broadband coverage and high sensitivity in the
iron K line region has now been achieved with \suzaku.
We then started a program aimed to observe with this satellite the brightest
(and best studied) BLRGs.
Here we present the X-ray view of the BLRG \treca\ obtained thanks to our \suzaku\ data 
and an archival long exposure with \xmm; Table~\ref{tab:src} 
summarizes the main global properties of the source.

%
\begin{table*}
\begin{minipage}[t]{2\columnwidth}
\caption{{\bf  Main global properties of \treca.}}
\label{tab:src}
\begin{center}
{
\begin{tabular}{c c c c c c c c c c}
 \hline \hline
  RA (J2000) & Dec (J2000) & Redshift & \pedix{N}{H,\,gal} & FWHM$\mbox{[H}\alpha\mbox{]}$ & $i$ & \pedix{M}{BH} & \pedix{R}{g} & \pedix{\log L}{bol} & $\log \pedix{L}{bol}/\pedix{L}{Edd}$ \\
   &  &  & [$10^{20}\,$\nh] & [km~s$^{-1}$] & [deg.] & [$10^{8}\pedix{M}{\sun}$] & [$10^{14}\,$cm] & [erg~s$^{-1}$] & \\
  \multicolumn{1}{c}{(1)} & \multicolumn{1}{c}{(2)} & \multicolumn{1}{c}{(3)} & \multicolumn{1}{c}{(4)} & \multicolumn{1}{c}{(5)} & \multicolumn{1}{c}{(6)} & \multicolumn{1}{c}{(7)} & \multicolumn{1}{c}{(8)} & \multicolumn{1}{c}{(9)} & \multicolumn{1}{c}{(10)} \\
 \hline
    $04$h$18$m$21.3$s & $+38$d$01$m$36$s & $0.0485$ & $29.1$ & $4800$  & $19$ & $36.0$ & $10.7$ & $45.56$ & $-2.11$ \\
     &  &  &  &  & $10-26$ & $1.5 - 2.4$ & $0.44 - 0.71$ & $44.40 - 45.26$ & $-$ \\
     &  &  &  &  &  & $1.8$ & $0.53$ & $44.68$ & $-1.69$ \\
 \hline
\end{tabular}
}
\end{center}       
{\footnotesize   
$(1)$~Right Ascension.
$(2)$~Declination.
$(3)$~Cosmological redshift.
$(4)$~Neutral Galactic absorption column density \citep{nh}.
$(5)$~Full Width Half Maximum of the optical line \citep{eracleous94,eracleous03}.
$(6)$~Inclination angle of the jet, from radio observations \citep{kadler08,eracleous00}; in the second row, the range
inferred by \citet{lewis05} is reported.
$(7)$~Black-hole mass: in the first row, from \citet{marchesini04}; in the second row, the range recently proposed by
\citet{chatt11}, which intermediate value is reported in the third row.
$(8)$~Schwarzschild radius; in each row, the value inferred from the corresponding black hole mass.
$(9)$~Bolometric luminosity: in the first row, estimated by \citet{marchesini04} from the optical luminosity; in the
second row, the range recently proposed by \citet{chatt11}, which intermediate value is reported in the third row.
$(10)$~Eddington ratio, from the black hole mass and bolometric luminosity reported in columns~(7) and (9),
respectively.
} \\
\end{minipage}
\end{table*}

The paper is organized as follows. 
The main properties of the source and previous X-ray results are presented
in Section~\ref{sect:blrg}.
Sections~\ref{sect:data} and \ref{sect:analysis} describe the 
X-ray observations and data analysis.
Our results are discussed in 
Section~\ref{sect:discuss}, and finally in Section~\ref{sect:summ} we 
summarize our work.
Throughout this paper, a
concordance cosmology with $\pedix{H}{0}=71\,$km s$^{-1}$ Mpc$^{-1}$,
$\pedix{\Omega}{\ensuremath{\Lambda}}=0.73$, and 
$\pedix{\Omega}{m}=0.27$ \citep{spergel03,spergel07} is adopted. 
The energy spectral index, $\alpha$, is defined such that
$\pedix{F}{\ensuremath{\nu}} \propto \apix{\nu}{-\ensuremath{\alpha}}$. 
The photon index is $\Gamma=\alpha+1$.


\section{The BLRG \treca}\label{sect:blrg}

\treca\ is a nearby ($z=0.0485$), well-studied BLRG with broad optical lines 
[FWHM$\mbox{(H}\alpha\mbox{)}\sim 4800\,$km s$^{-1}$; \citealt{eracleous03}].
It exhibits a Fanaroff-Riley II radio morphology, with a single-sided jet 
\citep{linfield84} showing superluminal motion \citep{vermeulen94}.
From the measured proper motion [$\mu=(1.54\pm 0.2)\arcsec\,$yr$^{-1}$] and 
the maximum angular size of the radio lobes 
\citep[$\theta=273\arcsec$;][]{nilsson93}, 
\citet{lewis05} inferred a range of $10\arcdeg<i<26\arcdeg$ 
($21\arcdeg<i<26\arcdeg$ unless the source is a giant radio galaxy) for the 
inclination angle of the jet.
A recent estimate of the jet inclination of $i\sim 19\arcdeg$ has been derived 
with a Very Long Baseline Array (VLBA) monitoring 
\citep[form the MOJAVE program; ][]{kadler08}.
This program revealed also the presence of a variety of 
components in the jet of \treca: a compact core, superluminal jet components, 
recollimation shocks, and regions of interaction between the jet and 
its surrounding medium.

From the relation with the bulge luminosity, \citet{marchesini04} estimated 
a black hole mass for \treca\ of 
$\pedix{M}{BH}=3.6\times 10^{9}\,\pedix{M}{\sun}$.
New estimates, derived from measurements of the $\mbox{H}\alpha$ width \citep{chatt11},
range from $1.5\times 10^{8}\,\pedix{M}{\sun}$ to $2.4\times
10^{8}\,\pedix{M}{\sun}$, more than a factor $10$ lower than the values derived by \citet{marchesini04}.
As suggested by \citet{chatt11}, this difference is probably due to the different extinction adopted in the two papers.

Several high-energy observatories targeted \treca.
Simultaneous \rxte\ and \xmm\ observations \citep{lewis05} revealed a rather 
flat continuum ($\Gamma\sim 1.63-1.75$, depending on the adopted model) and an extremely 
weak reflection component\footnote{The reflection fraction 
$R$ is defined as $R = \Omega/2\pi$, where $\Omega$ is the solid angle subtended by the 
Compton-thick
matter to the X-ray source; for a plane parallel slab, $R = 1$.}, $R \lesssim 0.3$.
To explain the broad residuals found in the iron K energy range, the authors 
assumed either reprocessed emission (i.e., reflection component and broad 
iron K emission line) from a truncated accretion disk, or
a partial covering absorber with high column density 
($\nhsym\sim 10^{23}\,$\nh);
although the two parametrizations were not distinguishable from the fitting point of view, the latter model, though
less complex, was disfavoured by the authors on physical grounds (mainly in view of the small inclination angle of the
jet).
Regardless of the continuum model used, a narrow \feka\ line arising in a 
distant reprocessor is required, with equivalent width (EW) $\sim 20-30\,$eV.

A \sax\ observation of \treca\ confirmed that a flat continuum is present ($\Gamma\sim 1.58$),
with only upper limits for the reflection component and the \feka\ line 
\citep[$\mbox{EW}\,<72\,$eV, $R < 0.3$;][]{grandi06}.
However, a re-analysis of the same \sax\ data presented in 
\citet{dadina07} found a steeper continuum 
($\Gamma\sim 1.75$) and a higher upper limit for the reflection parameter, 
$R \le 2.25$; the \sax\ data provides only a lower limit to the high-energy cutoff,
$\pedix{E}{c}\geqslant 82\,$keV.

The $6$-years of monitoring with \rxte\ recently presented in \citet{chatt11} highlights
the long time-scale variability of \treca.
The flux shows a range of variability
$\pedix{F}{2-10\kev}\sim 2\times 10^{-11}-8\times 10^{-11}\,$\flux\ (\ie, a factor of $\sim 4$), clearly correlated
with the variation in the X-ray photon index.
From their analysis of the variations in the X-ray continuum and \feka\ emission line, the authors conclude that the
iron line is generated within $90\,$light-days of the source of the X-ray continuum.
The latter can be either the corona or the base of the jet; in both cases, the \rxte\ data are consistent with the
standard paradigm of X-ray emission dominated by reprocessing of thermal photons produced in the accretion disk; no
strong contribution from the jet is observed.

A $10\,$ks \chandra\ exposure detected an excess of X-ray emission in $3$ radio knots out of the $4$ 
present along the single sided jet structure, as well as in its terminal point \citep{hogan11}.
The emission is ascribed by the authors to inverse Compton scattering off of
cosmic microwave background photons (IC/CMB).
Thus, depending on the adopted model for the jet structure from pc to kpc scales
(\eg, jet bending and/or deceleration, or neither of them), the combined 
\chandra\ and VLBI observations imply for the kpc-scale jet bulk Lorentz factor a range 
between $\sim 3.5$ and $7.2$.

In the $\gamma$-ray band, a $3\sigma$ \egret\ detection was reported for this source 
\citep{sguera05,hartman08}, as well as an association with a \fermi\ source
\citep{abdo10}, implying a broad-band SED reminiscent of a 
de-beamed blazar.
Comparing $\gamma$-ray and multiwavelength properties for a sample of
\fermi-detected and \fermi-undetected BLRGs, \citet{kataoka11} suggest that the
GeV emission from \treca\ is most likely dominated by the beamed radiation from
the nuclear region of the relativistic jet.


\section{X-ray observations}\label{sect:data}

\subsection{\suzaku} 

\suzaku\ observed \treca\ at the HXD \citep[Hard X-ray Detector,][]{hxd}
nominal pointing position on 
22 of August, 2008 
for a total exposure time of about $122\,$ks.
The log of the X-ray observations is reported in Table~\ref{tab:obslog}, first part.
We used the cleaned event files obtained from version~2 of
the \suzaku\ pipeline processing.  
Standard screening criteria were
used, namely, only events outside the South Atlantic Anomaly (SAA) as
well as with an Earth elevation angle (ELV) $ > 5\arcdeg$ 
were retained, and Earth day-time elevation angles (DYE\_ELV) $ > 20\arcdeg$. 
Furthermore, data within $256$~s of the SAA
passage were excluded and a cut-off rigidity of $ >6 \,\mathrm{GV}$
was adopted. 

XIS \citep[X-ray Imaging Spectrometer,][]{xis} data were selected in $3\times3$ and 
$5\times5$ editmodes using grades 
$0,2,3,4,6$, and cleaned for hot and flickering pixels.
The XIS spectra of \treca\ were extracted from circular regions 
centred on the sources of 2.9$\arcmin$ radius. 
Background spectra were extracted from two regions, each with the same area of the main target region, offset from the 
main target and avoiding the calibration sources.
The XIS response and ancillary response files were produced, using the latest calibration files
available, with the \textit{ftools} tasks \textit{xisrmfgen} and
\textit{xissimarfgen}, respectively.  

%
\begin{figure*}
  \centering
  \resizebox{0.33\hsize}{!}{\includegraphics{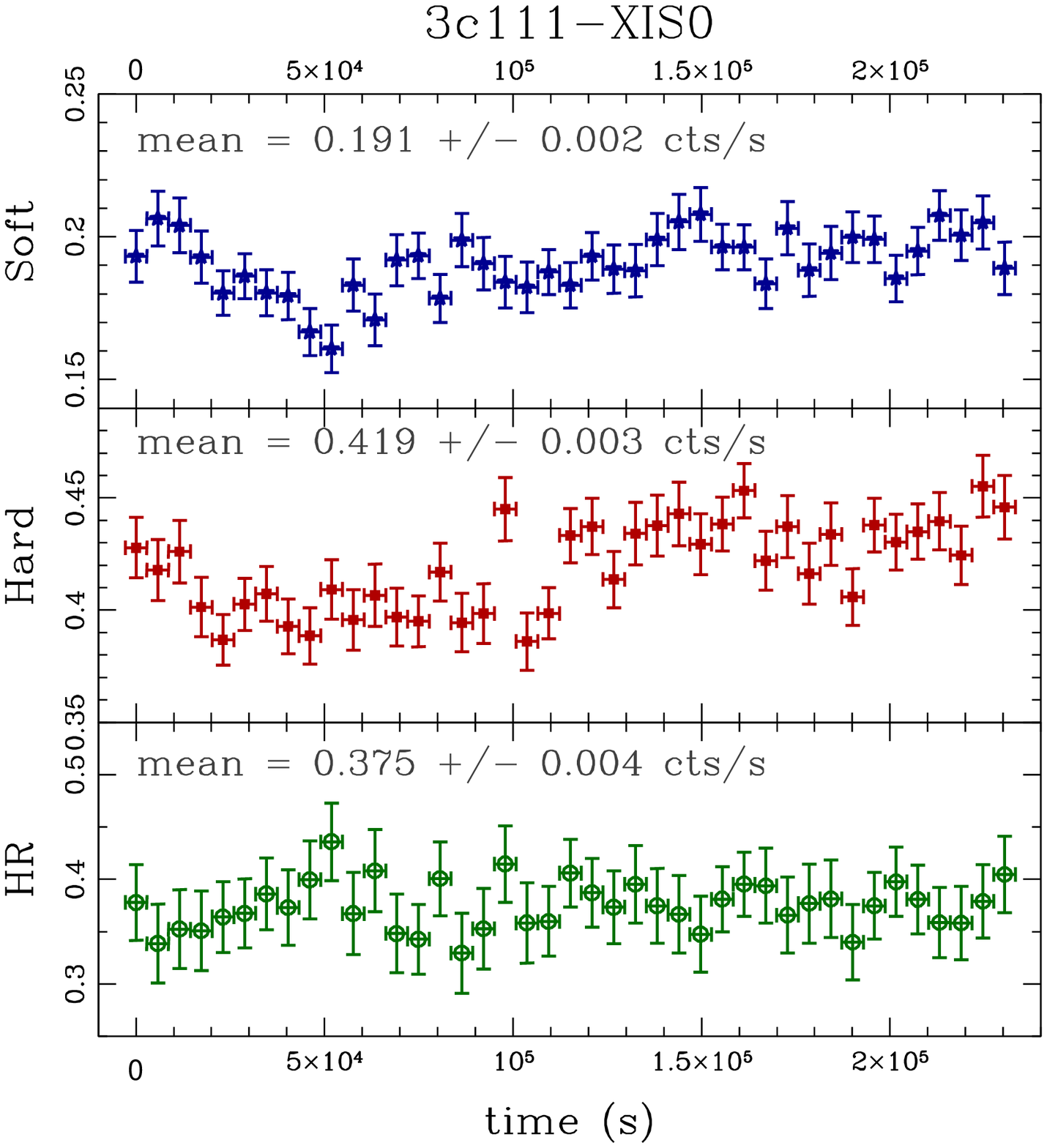}}
  \resizebox{0.33\hsize}{!}{\includegraphics{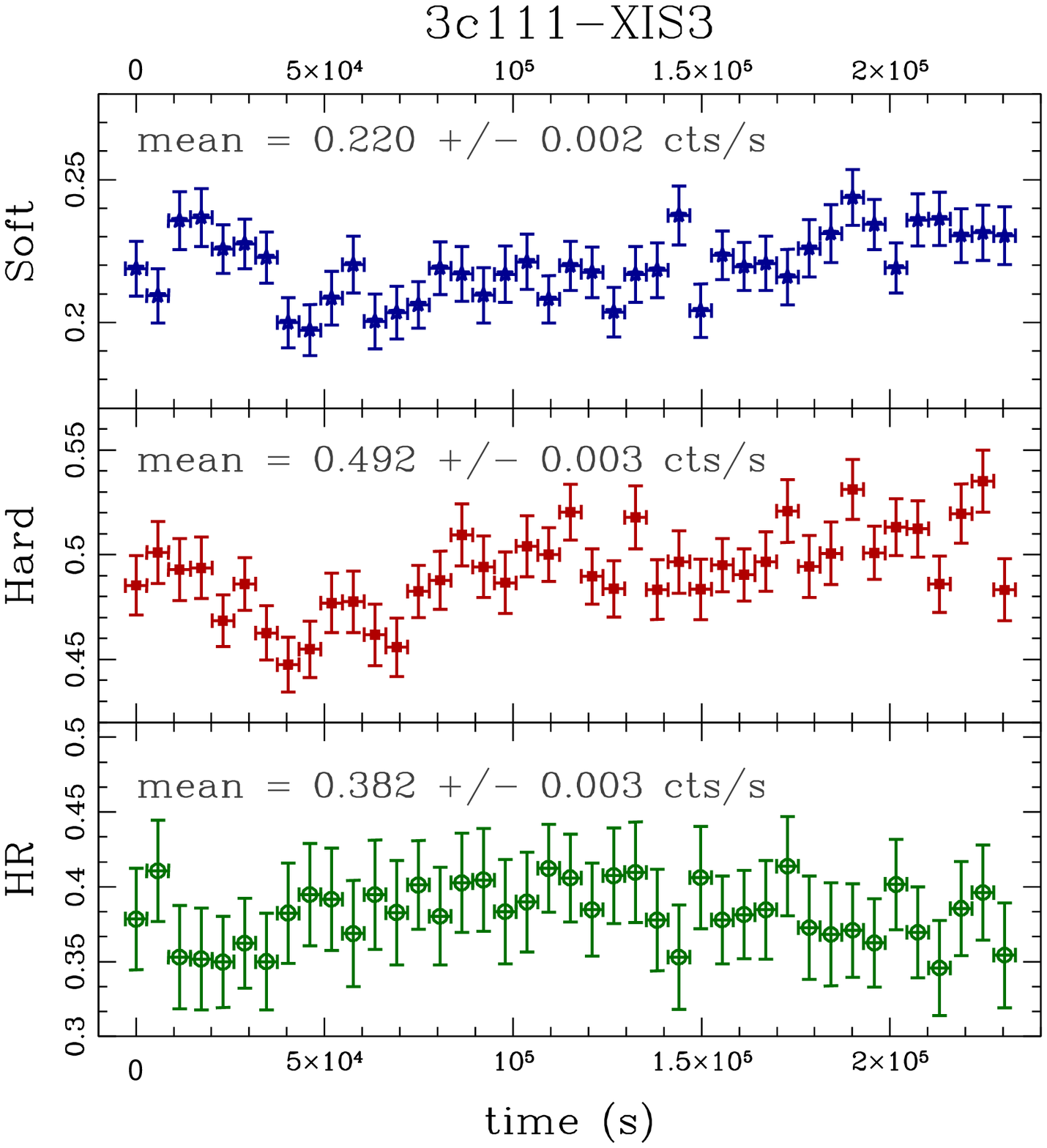}}
  \resizebox{0.33\hsize}{!}{\includegraphics{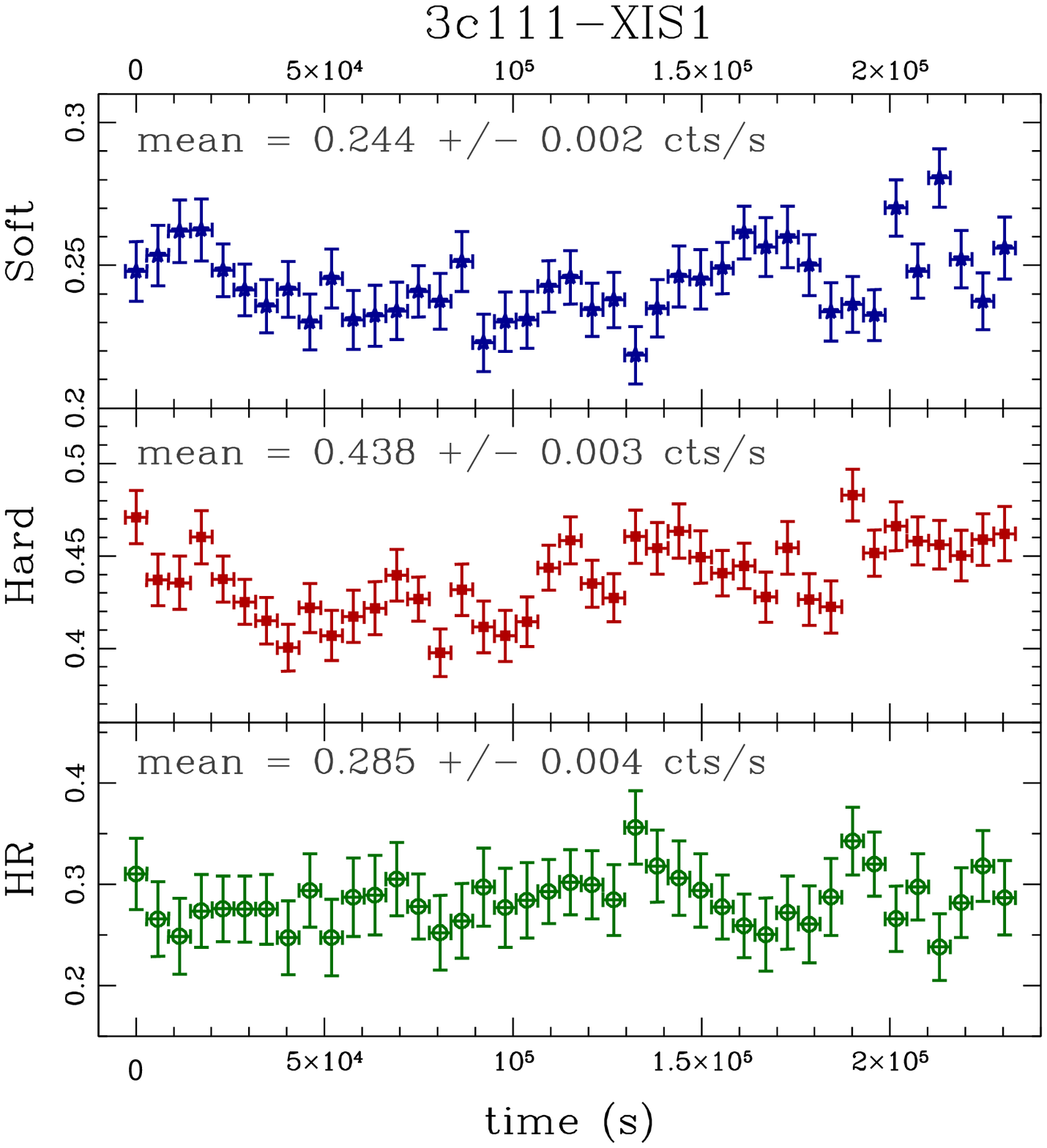}}
  \caption{Background-subtracted light curves of \treca\ from the \suzaku\ XIS0 ({\it left panels}), XIS3 
  ({\it middle panels}), and XIS1 ({\it right panels}) observations, with a $5760\,$sec bin.
  {\it Upper panels:} $0.5-2\,$keV; {\it middle panels:} $2-10\,$keV; {\it lower panels:} $HR=[\mbox{\sc
  rate}_{2-10}-\mbox{\sc rate}_{0.5-2}]/[\mbox{\sc rate}_{2-10}+\mbox{\sc rate}_{0.5-2}]$.
  In each panel, we report the mean rate [$HR$] for the whole observation.}
  \label{fig:3c111xislc}%
\end{figure*}

We tested time-variability within the \suzaku\ observation, generating source light curves in the soft and hard energy 
ranges with a binning time of $5760\,$sec, the orbital period of the satellite, to remove any possible modulation
related to the orbital condition.
Fig.~\ref{fig:3c111xislc} shows the soft and hard X-ray light curves obtained with the XISs, and their hardness
ratios.
The data from XIS0, XIS3, and XIS1 are plotted on separate panels, from {\it left} to {\it right}.
The source displays a hint of variability, with the $\chi^2$ probability of constancy less than 30\% and 5\% in the soft
and hard band, respectively.
However, the $HR$ light curves show a lack of significant spectral variability.
Therefore, in the following spectral analysis we considered the time-averaged data over the whole duration of the
observation.

After checking that the spectra were consistent, we coadded the two front 
illuminated (FI) CCD spectra\footnote{XIS0 and XIS3, found to be consistent 
within  $5$\% cross-normalization uncertainties}, along with the response files,
to maximize the signal-to-noise ratio.
The FI spectrum was then fitted jointly with the back illuminated (BI, the XIS1)
spectrum.

The net FI [BI] source spectra were rebinned to 1024 channels and then grouped 
with a minimum of $200$ [$100$] counts per bin.
Data were included from $0.6$ to $10$~keV [$8.5$~keV], observed frame; the region around the Si~K edge between 
$1.7$ and $1.9\,$keV were discarded in both the FI and BI XIS, due to
uncertainties in the calibration at this energy. 
The net background-subtracted source count rates for \treca\ were 
$0.609 \pm 0.002$, $0.712\pm 0.003$ and $0.676 \pm 0.003$~counts s$^{-1}$ for 
XIS0, XIS3 and XIS1 respectively, with a net exposure time after screening of 
$109$~ks.\\

For the HXD-PIN data reduction and analysis we followed the
\suzaku\ data reduction guide (the ABC guide\footnote{http://heasarc.gsfc.nasa.gov/docs/suzaku/analysis/abc/} 
Version 2).
We used the rev2 data, which include all four cluster units, and the
best background available, which account for non-X-ray background
\citep[NXB\footnote{ftp://legacy.gsfc.nasa.gov/suzaku/doc/hxd/suzakumemo-2008-03.pdf};][]{kokubun07}.

At the time of this analysis, two NXB files are available: background A or ``quick''
background and background D or ``tuned'' background. 
We adopted the latter, which is the latest release and which suffers 
lower systematic uncertainties of about 1.3\%, corresponding to about half 
uncertainty of the first release of the NXB.
We confirmed this choice as most reliable estimate of the NXB by comparing the 
background A or D light curve to the light curve obtained from the 
Earth occulted 
data (from Earth elevation angles ELV $< -5$): indeed, the Earth 
occulted data give a 
representation of the actual NXB rate, as this neither includes a 
contribution from 
the source nor from the cosmic X-ray background.

The source and background spectra were extracted within the common
good time interval and the source spectrum was corrected for the
detector deadtime.
The contribution of the diffuse cosmic X-ray background counts was
simulated using the spectral form of \citet{boldt87}, assuming the
response matrix for diffuse emission, and then added to the NXB.
With this choice of background, \treca\ is detected up to 
$70\,$keV at a level of $19$\% above the background. 
The net exposure time after screening was $102\,$ks.
The HXD-PIN spectrum was binned in order to have a signal-to-noise ratio
greater then $10$ in each bin, and the latest response file released by
the instrumental team was used.
The count rate in $14-70\,$keV is $0.081\pm0.002$~counts s$^{-1}$.
Assuming a single cutoff power-law component ($\Gamma\sim 1.6$ and 
$\pedix{E}{c}\sim 100\,$keV) this corresponds to a 
$\pedix{F}{14-70\kev}\sim 4\times 10^{-11}$\flux. 
In the spectral analysis, we used a cross-calibration constant of 
$1.18$ between the HXD and XIS spectra, 
as suggested by the \suzaku-HXD calibration team 
for observation at the HXD nominal pointing
position\footnote{http://heasarc.nasa.gov/docs/suzaku/analysis/watchout.html}.\\

Following the prescription\footnote{http://heasarc.gsfc.nasa.gov/docs/suzaku/analysis/hxd\_repro.html},
we reprocessed the GSO data from the unscreened events, using the new gain calibration as of August~2010, in order to 
apply the correct GSO gain history file.
Note that using the new GSO background instead of the one with the old gain calibration implies a loss of time of $\sim
33$\%.
As for the PIN, the GSO/NXB files was used in conjunction with the screened 
source events 
file to create a common good time interval.
The cosmic X-ray background, not included in the background event files, can 
be neglected, 
being less than $0.1$\% of the total background rate in the GSO.
Both source and background spectra were corrected for the detector deadtime, 
and the former was rebinned for the background subtraction, 
because the GSO background is created in $64$ bins.

\treca\ is marginally detected with the HXD/GSO. 
The background-subtracted GSO count rate in the $50-200\,$keV band is 
$0.06 \pm 0.03$~counts s$^{-1}$, corresponding to a $S/N \sim 1.3$ ($0.3$\% of 
the total), for a net exposure time of $68$~ks.

%
\begin{table*}
\begin{minipage}[t]{2\columnwidth}
\caption{{\bf Observation log.}}
\label{tab:obslog}
\begin{center}
{
\begin{tabular}{l l l l c c c}
 \hline \hline
  Satellite & Sequence N. & \multicolumn{1}{c}{Start (U.T.)} & \multicolumn{1}{c}{Stop (U.T.)} & Detector & Exposure & Count Rate \\
   &  &  &  &  & [ks] & [counts/s]\\
  \multicolumn{1}{c}{(1)} & \multicolumn{1}{c}{(2)} & \multicolumn{1}{c}{(3)} & \multicolumn{1}{c}{(4)} & \multicolumn{1}{c}{(5)} & \multicolumn{1}{c}{(6)} & \multicolumn{1}{c}{(7)} \\
 \hline
     \suzaku\ & 703034010 & 2008-08-22 09:37:00 & 2008-08-25 04:24:00 & XIS0 & $108.9$ & $0.609\pm0.002$\\
      &  &  &  & XIS3 & $108.9$ & $0.712\pm0.003$\\
      &  &  &  & XIS1 & $108.9$ & $0.676\pm0.003$\\
      &  &  &  &  PIN & $101.9$ & $0.081\pm0.002$\\
      &  &  &  &  GSO &  $68.1$ & $0.060\pm0.023$\\
 \hline
     \xmm\ & 0552180101 & 2009-02-15 19:41:57 & 2009-02-17 04:00:49 & pn & $59.1$ & $5.113\pm 0.009$\\
      &  & 2009-02-15 19:38:56 & 2009-02-17 04:00:34 & MOS2 & $60.9$ & $2.046\pm 0.006$\\
      &  & 2009-02-15 18:59:02 & 2009-02-17 04:10:44 & RGS1 & $83.3$ & $0.086\pm 0.001$\\
      &  & 2009-02-15 18:59:10 & 2009-02-17 04:10:44 & RGS2 & $83.2$ & $0.106\pm 0.001$\\
     \xmm\ & 0065940101 & 2001-03-14 12:56:20 & 2001-03-15 01:20:36 & RGS1 & $32.3$ & $0.132\pm 0.002$\\
      &  & 2001-03-14 12:56:20 & 2001-03-15 01:20:36 & RGS2 & $31.5$ & $0.162\pm 0.002$\\
 \hline
\end{tabular}
}
\end{center}       
{\footnotesize   
$(1)$~X-ray satellite.
$(2)$~Sequence number of the observation.
$(3)$~Starting date (in year-month-day) and time (in hh:mm:ss) of the observation.
$(4)$~End date (in year-month-day) and time (in hh:mm:ss) of the observation.
$(5)$~Detector on-board the satellite.
$(6)$~Exposure time after screening was applied to the data.
$(7)$~Net source count rate after screening and background subtraction, as 
observed in the $0.6-10\,$keV (XIS0 and XIS3), $0.6-8.5\,$keV (XIS1), $14-70\,$keV (PIN), $50-200\,$keV (GSO), 
$1-10\,$keV (pn, pattern $0$), $0.4-10\,$keV (MOS2, pattern $0$), and $0.65-2\,$keV (RGS) energy ranges.
} \\
\end{minipage}
\end{table*}

\subsection{\xmm}

\xmm\ observed \treca\ twice, in March~2001 for about a total of $45\,$ksec (Obs.~ID $0065940101$), and in February~2009
for about a total of $120\,$ks (Obs.~ID $0552180101$).
The analysis of the former observation is presented in \citet{lewis05}; the pn $2.4-10\,$keV light curve from the second
observation is included in the X-ray monitoring presented by \citet[their ``longlook data'']{chatt11}.
Here we present the analysis of the still
unpublished X-ray spectra obtained in the second pointing.

The observations were performed with the European Photon Imaging Camera (EPIC), the Optical Monitor (OM) and the
Reflection Grating Spectrometer (RGS).
The observation details are reported in Table~\ref{tab:obslog}, second part.
In this paper, we concentrate on the data in the X-ray band.

The three EPIC cameras \citep[pn, MOS1, and MOS2;][]{pn,mos} were operating in Small Window mode, with the thin filter
applied.
The \xmm\ data have been processed, and event lists for
the EPIC cameras were produced, using the Science Analysis Software (SAS version~10.0.2) with 
the most recent calibrations.

EPIC event files have been filtered for high-background time intervals, following the standard method consisting in
rejecting periods of high count rate at energies $>10\,$keV.
Given the X-ray brightness of the source, to avoid pile-up effects only events corresponding to pattern $0$ have been 
used \citep[see the \xmm\ Users' Handbook;][]{xmmhb}.
We have also generated the spectral response matrices at the source position using the SAS tasks {\it arfgen} and
{\it rmfgen}.

Source counts were extracted from a circular region of radius $25\arcsec$ and $30\arcsec$ for the MOS2 and pn cameras,
respectively.
For the pn camera, background counts were extracted from two source-free circular regions in the same chip of 
$50\arcsec$ radius each.
Background counts for MOS2 were extracted from similar regions but selected outside the central chip, symmetric with
respect to the source position.
Regarding the MOS1, in addition to two bad pixel columns near the source, a careful look to the images extracted 
in {\sc rawx,rawy} coordinates revealed the presence of two more dark columns falling in the centre of the source, not
visible using {\sc x,y} coordinates, strongly affecting the spectral shape.
Therefore, we decided to not use the data from the MOS1 camera, limiting the analysis of \xmm\ only to the MOS2 and pn
spectra.

According to the result of the SAS task {\it epatplot} applied to the cleaned event file, pn data are affected by
``X-ray loading'', \ie\ the inclusion of X-ray events in the offset map
calculation\footnote{http://xmm2.esac.esa.int/docs/documents/CAL-TN-0050-1-0.ps.gz}.
This phenomenon, observed in very bright sources, is a consequence of excessive count rate, and occurs at lower source
count rate than the  pile-up threshold.
The effects are a systematic energy shift to low energy and pattern migration from doubles to single events.
As the pn offset maps are calculated on-board at the start of each exposure, for this camera the effects are hard to be
reduced, therefore the pn data are more easily prone to this problem than the MOS one.
At present, a reliable method to correct for this effect is not available.
We then decided to discard the pn data in the soft energy range, more affected by the problem.
The spectral shape of the pn spectrum above $1\,$keV is consistent with the MOS2 one, and with the \suzaku\ spectrum; 
this confirms that the selected band is not affected by the X-ray loading problem.
Therefore, during the analysis we considered the energy range between $0.4$  [$1$] and $10\,$keV for the MOS2 and the 
pn, respectively.

As previously performed for \suzaku, we checked pn and MOS2 data for variability during the observation using the SAS 
task {\it lcplot}.
No bin shows significant deviation from the mean value, allowing us to perform the spectral analysis over the 
time-averaged spectra.
Both pn and MOS2 spectra were grouped to have at least $200$ counts in each energy bin.

The RGS \citep{rgs} data have been reduced
using the standard SAS task {\it rgsproc}, and the most recent calibration 
files; after filtering out the high-background time intervals, the total 
exposure times are $\sim 83\,$ks for both RGS1 and RGS2. 
The RGS1 and RGS2 spectra were binned
at twice the instrument resolution, $\Delta\lambda=0.2\,$\AA.

\subsection{\swift\ BAT}

The averaged BAT spectrum of this radiogalaxy was obtained from the $58$-month 
survey archive (SWIFT~J$0418.3+3800$). 
The data reduction and extraction procedure of the 8-channel spectrum is
described in \citet{swift58}. 
To fit the preprocessed, background-subtracted BAT spectra, we used the latest 
calibration response {\tt diagonal.rsp} as of December 2010. 

\treca\ was detected in the $15-100\,$keV band with a count rate 
$(1.02 \pm 0.03) \times 10^{-3}$~counts s$^{-1}$, which correspond to a 
$14-195\,$keV flux of $1.2 \pm 0.1 \times10^{-10}\,$\flux\ \citep{swift58}.

Fitting the averaged BAT spectrum with the HXD best-fit model
($\Gamma\sim1.6$ and $\pedix{E}{c}\sim 100\,$keV), we found 
$\pedap{F}{14-70\kev}{BAT}\sim 7\times 10^{-11}\,$\flux, while 
$\pedap{F}{14-70\kev}{PIN}\sim 4\times 10^{-11}\,$\flux (see Table~\ref{tab:111fllum}).
The comparison between the BAT and the PIN spectra
suggests a possible hard X-ray variability on longer timescales: fitting the 
two spectra together and allowing to
vary the relative normalizations, we found that we need  
the cross-normalization between the two instruments
to be $\sim 1.6$, well above the uncertainties in the relative 
normalizations.
The $58$-month BAT lightcurve\footnote{http://swift.gsfc.nasa.gov/docs/swift/results/bs58mon/mosaic\_crab\_lc/\\
BAT\_58m\_crabweighted\_monthly\_SWIFT\_J0418.3+3800.lc.gz}, provided with
the catalogue, suggests the presence of hard X-ray variability and that the
\suzaku\ observation took place during an upswing
after reaching a minimum of the $14-195\,$keV flux.
We caution that the \swift\ spectrum used here is the averaged spectrum.

\section{X-ray Spectral Analysis}\label{sect:analysis}

All spectral fits to the X-ray data were performed using 
{\tt XSPEC} v.12.6.0q.
The significance of adding free parameters to the
model was evaluated with the F-test, with associated probability
\pedix{P}{F}\footnote{But see the caveats in using the F-test to 
measure the significance of narrow lines described in \citet{protassov02}.}. 
All uncertainties are quoted at the 90\% confidence level for one parameter of interest ($\Delta\apix{\chi}{2}=2.71$).
Unless otherwise stated, the figures are in the rest-frame of \treca, and fit parameters are quoted in the same frame.
All the models discussed in the following assume Galactic absorption 
with a column density of $\pedix{N}{H, Gal}\sim 2.91\times10^{21}\,$\nh\ 
\citep[see Table~\ref{tab:src};][]{nh}.
It is worth noting that for \treca\ the estimate of the total Galactic 
hydrogen column density is not trivial: due to the presence of a dense 
molecular cloud along the line of sight, its value can as high as 
$\pedix{N}{H, Gal}\sim1.3\times10^{22}\,$\nh\ 
\citep[and references therein]{lewis05}, and it is expected to vary 
by several times $10^{21}\,$\nh.
Therefore we allowed the local absorption to vary up to  $1.3\times10^{22}\,$\nh; for this absorber 
we used the {\tt phabs} model in {\tt XSPEC}, adopting
cross-sections and abundances of \citet{wilms00}.

\subsection{\suzaku}\label{sect:an3c111suz}

\begin{figure}
  \centering
  \resizebox{\hsize}{!}{\includegraphics[angle=270]{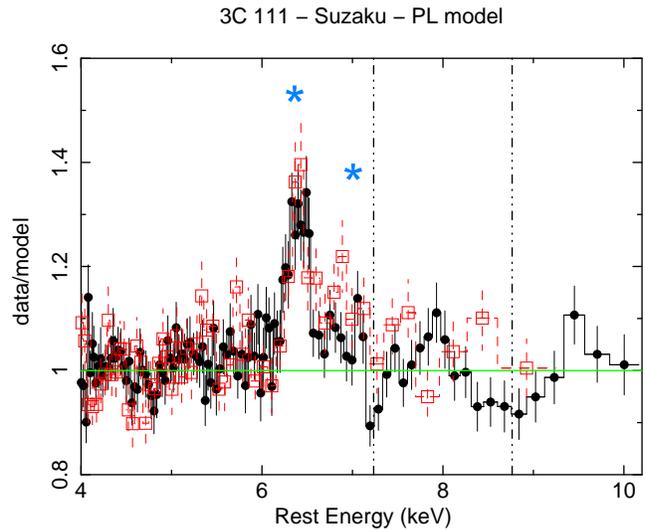}}
  \caption{Data-to-model ratio for the \suzaku\ spectra (red open squares, 
XIS1; black filled circles and line, FI XIS).
    The model is an absorbed power-law component ($\Gamma \sim 1.61$) 
fitted ignoring the $5-7.5\,$keV energy band, where the iron K complex is
expected.
    For graphical purposes, the data have been binned to have a significance of $17\sigma$.
    The positions of the iron K lines (\feka\ at $6.4\,$keV and \fekb\ at 
$7.06\,$keV rest frame) are marked with a blue star.
    The vertical lines represent the centroids of the detected absorption lines 
in the FI XIS data 
    \citep[see][]{tombesi10}.}
  \label{fig:3c111ra}%
\end{figure}

The medium-energy spectrum of \treca\ confirms the 
presence of strong iron K emission detected in previous observations 
\citep{lewis05,grandi06,dadina07}.
We first fitted the XIS and PIN data (between $0.6$ and $70\,$keV) with a 
single absorbed power-law component,
excluding the energy range where the iron K complex is expected ($5-8\,$keV).
After including the $5-8\,$keV data, this model clearly provides a poor fit 
[$\chidof=1046.70/743$; $\Gamma = 1.61 \pm 0.01$ and 
$\nhsym = (1.20 \pm 0.02) \times 10^{22}\,$\nh]; furthermore, large-amplitude 
residuals are present both in emission and in absorption (see  
Fig.~\ref{fig:3c111ra}).
No additional intrinsic \nhsym\ is required ($\dchidof=0.06/1$); 
however,  taking into account the strong variability of \pedix{N}{H, Gal} for
\treca,  we cannot exclude that some of  the observed absorption  is at the source redshift.

%

The addition of a Gaussian line to the simple absorbed power-law component 
results in a 
large improvement in the fit ($\dchidof = 244.5/3$).
The profile of the \feka\ line ($E = 6.40 \pm 0.02\,$keV, 
$\sigma=94\errUD{31}{22}\,$eV; $F=2.36\errUD{0.46}{0.26}\times 10^{-5}\,$\normGAUSS,
$\mbox{EW}=94\errUD{13}{18}\,$eV) is 
marginally resolved, suggesting a possible contribution from the external disk and/or the BLR.
However, constraining the iron K line to be narrow ($10\,$eV) does not leave strong residuals 
which could be due to an additional broad component.
The strength of the \feka\ line, coupled with the hard photon index 
($1.59 \pm 0.01$) could suggest the presence of a reflection component.
However, adding a reflection component \citep[parametrized using the 
{\tt pexrav} model in {\tt XSPEC};][]{pexrav} to the absorbed cutoff power 
law plus iron line, we obtained a weak reflection fraction, 
$R = 0.16 \errUD{0.11}{0.09}$ ($\chidof=782.7/738$).
During the fit, the abundance was fixed to the Solar one; the photon 
indices of the cutoff power law and the reflection component, as well as their energy cutoff, were set to be equal.
Allowing the high energy cutoff to vary we can set
only a lower limit ($\pedix{E}{c}>75\,$keV). 
The inclination angle 
\citep[allowed to vary between $10\arcdeg$ and $26\arcdeg$;][]{lewis05}, 
is also not constrained by the present data.
We therefore decided to fix both parameters (see below).

Although the energy centroid of the \feka\ line suggests a low
ionization reflector, we tested for a possible
ionized reflector. 
We used an updated model for the Compton
reflection off an optically thick photoionized slab of gas, which
includes the iron K emission line \citep[{\tt reflionx} model in {\tt XSPEC};][]{reflionx1,reflionx2}; as expected the
ionization of the reflector is found to be low, $\xi < 13.9\,$erg~cm~s$^{-1}$, being the
ionization parameter mainly driven by the energy centroid of the \feka\ line.

%
\begin{table}
\begin{minipage}[t]{\columnwidth}
\caption{{\bf X-ray fluxes and luminosities.}}
\label{tab:111fllum}
\begin{center}
{\footnotesize
\begin{tabular}{c c c c}
 \hline \hline
   Satellite & \multicolumn{3}{c}{Flux$^{a}$} \\
   & \multicolumn{3}{c}{[$10^{-12}\,$\flux]} \\
   \cline{2-4}
   & $0.5-2\,$keV & $2-10\,$keV & $14-70\,$keV \\
 \hline
   \suzaku\ & $2.55\errUD{0.05}{0.06}$ & $19.70\errUD{0.52}{0.48}$ & $40.21\errUD{1.99}{1.91}$ \\
   \xmm\ & $6.83\errUD{0.07}{0.11}$ & $48.32\errUD{1.10}{0.87}$ & $-$ \\
   \swift\ & $-$ & $-$ & $71.19\pm 2.45$ \\
 \hline \hline
   Satellite & \multicolumn{3}{c}{Luminosity$^{b}$} \\
   & \multicolumn{3}{c}{[$10^{43}\,$\lum]} \\
   \cline{2-4}
   & $0.5-2\,$keV & $2-10\,$keV & $14-70\,$keV \\
 \hline
   \suzaku\ & $5.44\errUD{0.16}{0.15}$ & $11.03\errUD{0.28}{0.27}$ & $21.69\errUD{1.07}{1.03}$ \\
   \xmm\ & $12.82\errUD{0.23}{0.11}$ & $27.54\errUD{0.62}{0.48}$ & $-$ \\
   \swift\ & $-$ & $-$ & $39.35\pm 1.36$  \\
 \hline
\end{tabular}
}
\end{center}       
{\footnotesize  Errors are quoted at the 90\% confidence level for 1 parameter 
of interest ($\Delta\apix{\chi}{2}=2.71$). 
$^{a}\,$Observed fluxes.
$^{b}\,$Intrinsic luminosities, corrected for local absorption, as well as for the effects of the 
photoionized gas. 
} \\
\end{minipage}
\end{table}

As the reflection component is weak and poorly constrained, we bounded its 
normalization to that of 
the emission line: to fit the two components in a consistent way, the ratio of 
their normalizations was assumed to be the expected 
one\footnote{http://www.jca.umbc.edu/$\sim$george/html/science/seyferts/fgfplus.html} for a 
face-on slab illuminated by a flat continuum ($\Gamma\sim1.6$), 
i.e. $\pedix{N}{gauss}/\pedix{N}{pexrav}\sim 1.2 \times 10^{-2}$ 
\citep{george91,nandra94,matt96,matt97}.
The $R$ parameter was fixed to $1$, the inclination angle to 
$19\arcdeg$ \citep{kadler08}, and the energy cutoff of the primary power law and of the reflection component to
$100\,$keV, a value consistent with the lower limit previously found.
Higher values of $\pedix{E}{c}$ imply a slight decrease in the quality of the fit.
We also included a neutral \fekb\ line (at an energy fixed to 
$7.06\,$keV) with a flux of $13.5$\% of the \feka\ line \citep{palmeri03,yaqoob10} and the same width.

The baseline model then consists of: a rather flat cutoff power law 
($\Gamma = 1.61\pm 0.01$), and a reflection continuum$+$two narrow iron K 
lines.
The width of the line, allowed to vary during the fit, is $\sigma = 70\errUD{27}{28}\,$eV.
The strength of the reflection, estimated as the ratio of the {\tt pexrav} and the
direct power-law normalizations, is 
$\pedix{N}{pexrav}/\pedix{N}{cutoffpl}=0.35\pm 0.06$.
The column density is 
$\nhsym=(1.20 \pm 0.02) \times 10^{22}\,$\nh.
This model provides an acceptable description for the 
broadband X-ray spectrum of \treca\ ($\chidof=793.8/740$).

%
\begin{figure}
  \centering
  \resizebox{\hsize}{!}{\includegraphics[angle=270]{f3.ps}}
  \caption{Data-to-model ratio for the \suzaku\ spectra below $1.5\,$keV; red open squares, 
XIS1; black filled circles and line, FI XIS.
    For graphical purposes, the data have been binned to have a significance of $10\sigma$.
    The adopted model is the baseline model (an absorbed  cutoff power law and 
a reflection component, with superimposed two narrow Gaussian lines, the 
\feka\ at $6.4\,$keV and the \fekb\ at $7.06\,$keV rest frame).
    The centroid of the emission line when one more Gaussian is included in the model is marked with a blue star 
    (emission line at $E\sim0.89\,$keV).
    }
  \label{fig:3c111le}%
\end{figure}
%

%
\begin{figure}
  \centering
  \resizebox{\hsize}{!}{\includegraphics[angle=270]{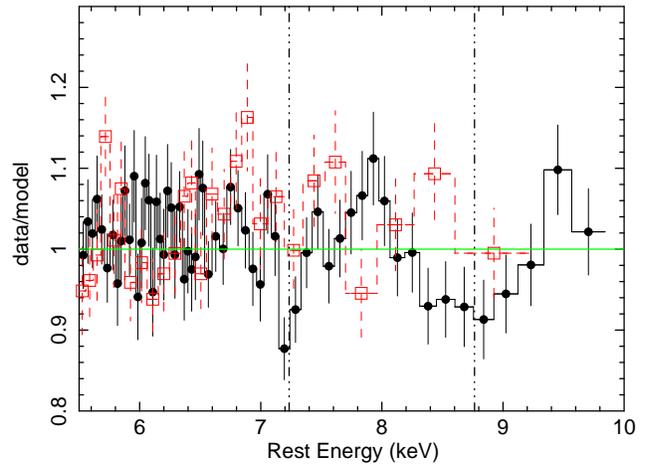}}
  \caption{Data-to-model ratio for the \suzaku\ spectra above $5.5\,$keV; red open squares, 
XIS1; black filled circles and line, FI XIS.
    For graphical purposes, the data have been binned to have a significance of $17\sigma$.
    The adopted model is the baseline model (an absorbed  cutoff power law and 
a reflection component, with superimposed two narrow Gaussian lines, the 
\feka\ at $6.4\,$keV and the \fekb\ at $7.06\,$keV rest frame).
    The centroids of the absorption lines
when two more Gaussians are included in the model are marked with two vertical lines (absorption line at
$E\sim 7.2\,$keV and $E\sim 8.8\,$keV).
    }
  \label{fig:3c111fe}%
\end{figure}

However, some residuals suggest the presence of lines in emission, around $0.9\,$keV,
and in absorption, between $6$ and $9\,$keV (see Fig.~\ref{fig:3c111le} and Fig.~\ref{fig:3c111fe}).
To account for this residuals we first added to the baseline model $3$ 
Gaussian lines ($2$ absorption and one emission lines). 
Including a narrow emission Gaussian component  ($\sigma$ fixed to $10\,$eV)
at low energies, the improvement in the 
fit is significant ($\dchidof=10.2/2$, considering only $736$ XIS bins between 
$0.6$ and $10\,$keV); the energy is 
$E=0.89 \pm 0.01\,$keV, with $\mbox{EW}=32\errUD{15}{10}\,$eV.
Tentatively, the line can be identified with the \oviii~RRC or the \neix~triplet lines.
The two absorption lines are detected at the rest-frame energies of $E=7.23\errUD{0.06}{0.07}\,$keV 
($\sigma$ fixed to $50\,$eV; $\mbox{EW}=-15.2\errUD{10.8}{9.9}\,$eV) and $E=8.76\pm 0.20\,$keV 
($\mbox{EW}=-48.9\errUD{28.5}{36.2}\,$eV), with a $\dchidof=9.5/2$ and $10.9/3$,  respectively.
We refer to the Monte Carlo simulations presented in \citet{tombesi10} for a more robust 
estimate of the significance of the high-energy absorption lines in this \suzaku\ observation of \treca.
We note that in their analysis the authors considered only the FI data, and adopted for the $3.5-10.5\,$keV continuum a 
single power law, resulting in a flatter photon index $\Gamma\sim 1.47$
and a higher continuum model above $\sim 7\,$keV.
However, given the limited energy band considered, the effect of this continuum model
does not strongly 
affect the detection significance of the absorption lines.
If we consider only FI data, as the XIS1 is less reliable in the iron K energy range, the significance of the 
absorption lines is: $\dchidof=15.5/2$ and $15.5/3$ for the line at $E\sim7.23\,$keV and $E\sim8.76\,$keV, 
respectively.

A detailed study of the absorption features and their interpretation in terms of
ultra-fast outflows related with the accretion disk is reported in 
\citet{tombesi10}. 
Briefly, the most plausible association for the first absorption line is the resonant \ion{Fe}{xxvi} 
\Lya\ transition at $6.97\,$keV.
If the line is identified with this atomic transition, we have to assume a velocity shifts of 
$\pedix{v}{out}\sim -0.04\,c$.
Concerning the second feature, it is likely a blend of different lines of the \ion{Fe}{xxvi}
Lyman series (namely, \Ly{\beta}, \Ly{\gamma}, and \Ly{\delta}).
In a more physically consistent modelling of the absorption feature, the 
absorber is parametrized using a grid of 
photoionized absorbers generated by the 
{\tt XSTAR}\footnote{http://heasarc.gsfc.nasa.gov/docs/software/xstar/xstar.html} 
photoionization code \citep{xstar1,xstar2}. 
The photoionized gas is assumed to be fully covering the X-ray source with turbulence velocity 
$\pedix{v}{turb}= 500\,$km~s$^{-1}$, and illuminated by a power-law continuum
from $0.0136$ to $13.6\,$keV with photon index $\Gamma=2$.
It is important to note that a single photoionized absorber can account for both lines.
The results of the fit are: high ionization state, $\log \xi = 5.3\errUD{0.6}{1.2}\,$erg~cm~s$^{-1}$, with 
$\pedap{N}{H}{ion}\sim 6.97\times 10^{23}\,$\nh\ ($\pedap{N}{H}{ion} > 1.45\times 10^{22}\,$\nh\ at the $90$\% level),
and blueshift velocity $\pedix{v}{out}= 0.038\pm 0.006\,c$ ($\chi^2=768.2$ for $735$ degree of freedom).
We note that the high ionization state implied by $\log \xi \sim 5.3\,$erg~cm~s$^{-1}$ does not introduce a significant
curvature in the primary continuum below $\sim 6\,$keV.
The outflow velocity is somewhat similar to the source cosmological redshift, $z = 0.0485$, and thus could suggests a
contamination due to local or intermediate redshift absorbers \citep[\eg,][]{mckernan03,mckernan04,mckernan05}.
However, this scenario is ruled out by the high column density implied for this absorber. 
We refer the reader to the discussion in \citet[sect.~5.1]{tombesi10}, where the authors investigate in details this
possibility and conclude that it is not feasible.
The addition of the photoionized absorber 
improves the fit of $\dchidof=13.8/3$.

\begin{figure}
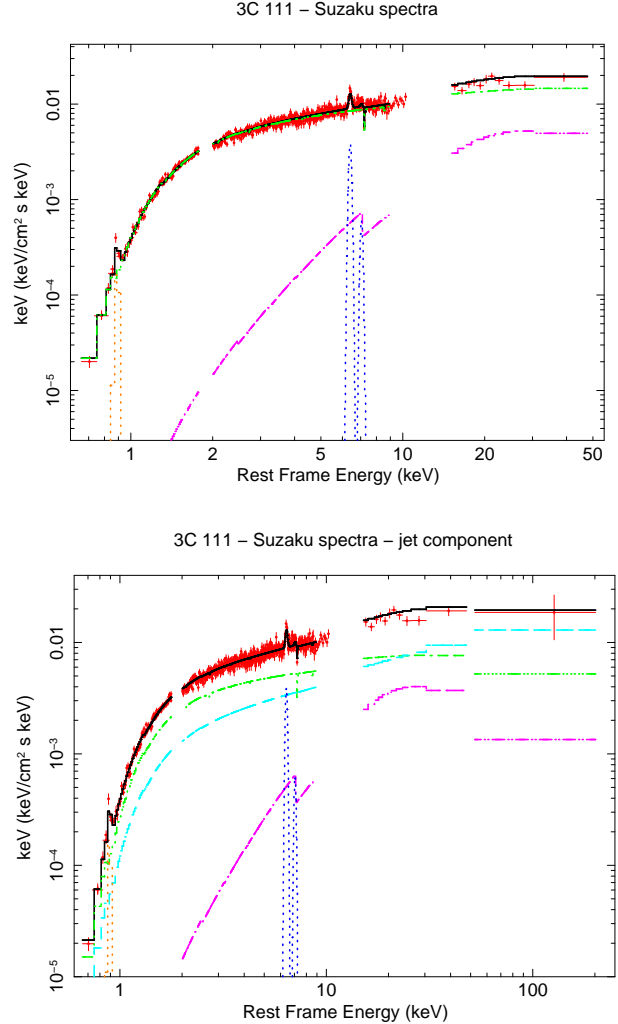

  \centering
  \resizebox{0.95\hsize}{!}{\includegraphics[angle=270]{f5a.ps}}
\vspace{0.5cm}\\
  \resizebox{0.95\hsize}{!}{\includegraphics[angle=270]{f5b.ps}}
  \caption{\suzaku\ XIS and HXD spectra of \treca, binned to have a significance of $3\sigma$.
  In both panels, the black lines represent the total best-fit model; the underlying continuum
  is modelled
  with a cutoff power-law component (green dash-dot-dotted lines) plus a weak reflection component (magenta
  dash-dotted lines).
  The iron K line complex is composed by a \feka\ and a \fekb\ features (blue dotted lines).
  In addition, an X-ray emission line at $\sim0.9\,$keV is also shown (orange dotted lines).
  In the {\it bottom panel} we show the model when a component due to the jet is considered (light blue line); also
  reported is the detection in the GSO, extending the spectrum up to $200\,$keV.
    }
  \label{fig:3c111eeuf}%
\end{figure}

Observed fluxes and intrinsic luminosities are summarized in Table~\ref{tab:111fllum};
the unfolded best fit to the data is shown in Fig.~\ref{fig:3c111eeuf} ({\it upper panel}).
The best-fit parameters are reported in the first part of Table~\ref{tab:111bfpar}; note that the 
inclusion of the warm absorber does not change significantly the continuum photon index or the emission lines
parameters.

%

As a final test, taking into account the low reflection component and the flat 
continuum observed, we checked the presence of a second continuum emission due
to the jet.
To this end, we considered also the GSO detection of \treca, that extends the source spectral energy distribution 
(SED) up to $\sim 200\,$keV.
Adding to the best fit model a second power law, with a flatter photon index 
($\pedix{\Gamma}{jet}=1.51\pm 0.03$), 
we obtained for the primary cutoff power law a photon index in better agreement with values 
expected from disk emission, $\Gamma=1.70\errUD{0.06}{0.02}$.
We fixed the parameters of the lines, as well as the column density, the ionization state, and the velocity 
of the outflow, to the best-fit values previously found.
This model provides a good description of the broadband X-ray spectrum of \treca, $\chidof=846.6/768$ (see
Fig.~\ref{fig:3c111eeuf}, {\it bottom panel}).
Despite the inclusion of the GSO data, the improvement in the fit due to the
addition of this component is only $\dchidof = 3.1/2$ (F-test probability $\pedix{P}{F}=75.6$\%).
We note that the jet dominates only above $10\,$keV, thus the possibility to accommodate a jet component in the
best-fit model is strongly 
related to the inclusion of the GSO detection in the fit.
Moreover, the relative normalizations of the two power laws are degenerate, making it difficult to spectrally
distinguish the true jet component.

%
\begin{table*}
\begin{minipage}[t]{2\columnwidth}
\caption{{\bf Best-fit parameters for the the \suzaku\ (XIS+PIN, $0.6-70\,$keV) and the \xmm\ [MOS2+pn, $0.4(1)-10\,$keV] spectra of \treca.}}
\label{tab:111bfpar}
\begin{center}
{\footnotesize
\begin{tabular}{c c c c c c c c c c c c}
 \hline \hline
   \multicolumn{3}{c}{Warm absorber$^{a}$} & & \multicolumn{3}{c}{Direct \& reflected continua} & & \multicolumn{3}{c}{Emission lines$^{b}$} & \\
   \cline{1-3} \cline{5-7} \cline{9-11}
  \pedap{N}{H}{ion} & $\log\xi$ & \pedSM{v}{out} & & $\Gamma$ & \pedSM{E}{c} & $R$ & & \pedSM{E}{g} & $\sigma$ & EW & \chidof \\
   $\mbox{[$10^{22}\,$\nh]}$ & [erg~cm~s$^{-1}$] & [$c$] &  &  & [keV] &  &  & [keV] & [eV] & [eV] & \\
   (1) & (2) & (3) & & (4) & (5) & (6) & & (7) & (8) & (9) & (10) \\
 \hline
   \multicolumn{12}{c}{\suzaku}\\
 \hline
    $69.68\errUD{22.59}{68.23}$ & $5.26\errUD{0.59}{1.16}$ & $0.038\pm 0.006$ & & $1.61\errUD{0.02}{0.01}$ & $100^{\mydag}$ & $0.35\pm 0.06$ & & $6.40\pm0.02$ & $74\pm 27$ & $75\pm 13$ & $772.5/735$ \\
     &  &  &  &  &  &  &  & $7.06^{\mydag}$ & $74$ & $12$ &  \\
     &  &  &  &  &  &  &  & $0.89\pm0.01$ & $10^{\mydag}$ & $33\errUD{13}{12}$ &  \\
 \hline
   \multicolumn{12}{c}{\xmm$^{c}$}\\
 \hline
   $-$ & $-$ & $-$ & & $1.58\pm 0.01$ & $100^{\mydag}$ & $0.19\errUD{0.05}{0.04}$ & & $6.41\pm 0.03$ & $89\errUD{53}{51}$ & $38.2\errUD{10.7}{9.2}$ & $1219.7/1144$ \\
    &  &  &  &  &  &  &  & $7.06^{\mydag}$ & $89$ & $6.0$ &  \\
 \hline
\end{tabular}
}
\end{center}       
{\footnotesize  The adopted model is a cutoff power law plus a reflection component, covered by a
photoionized absorber, with the addition of the iron K complex; for the \suzaku\ data, a low-energy narrow emission line
is also included.
Local \nhsym\ are $(1.21\pm 0.02)\times 10^{22}\,$\nh\ and $(1.04\pm 0.01)\times 10^{22}\,$\nh\ for the \suzaku\ 
and \xmm\ spectra, respectively.
$^{a}\,$Parametrized with an {\tt XSTAR} component obtained assuming a turbulence velocity 
$\pedix{v}{turb}= 500\,$km~s$^{-1}$.
$^{b}\,$The ratio of the \feka\ line and the reflection component normalizations was fixed to $0.012$ (see text); 
the \fekb\ and \feka\ ratio line was fixed to $0.135$; the \fekb\ line width was coupled
to the \feka\ line width.
$^{c}\,$For the \xmm\ spectra, the highly ionized absorber is not required by the data; the inclusion of this component
implies:
$\pedap{N}{H}{ion}=0.21\errUD{0.10}{0.08}\times 10^{22}\,$\nh, $\log\xi=2.80\errUD{0.22}{0.20}\,$erg~cm~s$^{-1}$, and
$\pedSM{v}{out}=0.0010 \pm 0.008\,c$.
$^{\mydag}\,$Parameter fixed. 

$(1)$~\nhsym\ of the ionized absorber.
$(2)$~Ionization parameter.
$(3)$~Outflow velocity.
$(4)$~Cutoff power-law and reflection component photon index.
$(5)$~Cutoff energy.
$(6)$~Reflection fraction (calculated as the ratio of the reflection component and the direct continuum).
$(7)$~Line energy.
$(8)$~Line width.
$(9)$~Line equivalent width.
$(10)$~$\chi^2$ and number of degree of freedom.
} \\
\end{minipage}
\end{table*}

\subsection{\xmm}\label{sect:an3c111xmm}

\begin{figure}
  \centering
  \resizebox{\hsize}{!}{\includegraphics[angle=270]{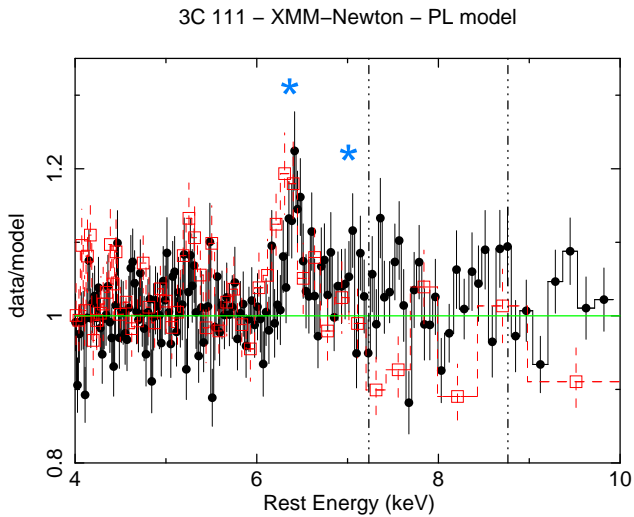}}
  \caption{Data-to-model ratio for the \xmm\ spectra (red open squares, 
MOS2; black filled circles and line, pn).
    The model is an absorbed power-law component ($\Gamma \sim 1.60$) 
fitted ignoring the $5-7.5\,$keV energy band, where the iron K complex is
expected.
    The blue stars and the vertical lines mark the position of the features detected in the XIS data, as in
    Fig.~\ref{fig:3c111ra}.
    For demonstration purposes, the data have been binned to have a significance of $20\sigma$.}
  \label{fig:xmm3c111ra}%
\end{figure}

The $0.6-10\,$keV \xmm\ emission shape of \treca\ is in good agreement with the best fit for the broad band continuum of
\suzaku, both in the broad-band slope ($\Gamma\sim 1.60$, $\nhsym \sim 1.04\times 10^{22}\,$\nh) and in the detection of
a strong \feka\ emission line (see Fig.~\ref{fig:xmm3c111ra}).
The main difference is in the observed flux, a factor $\sim 2.5$ higher than the \suzaku\ flux, see
Table~\ref{tab:111fllum}.
This is not surprising when the \rxte\ long-term light curve is considered \citep[see fig.~1 in][]{chatt11}: the
\suzaku\ observation took place during a recovery of the emission after an historical low state, while the 
\xmm\ data corresponds to the average $2-10\,$keV flux level.

%

Taking into account the \suzaku\ results and the evident presence of iron K emission in the \xmm\ spectra,
we then investigated the presence of an emission line plus a reflection component, the latter parametrized again with 
the {\tt pexrav} model
in {\tt XSPEC}, with abundance fixed to the Solar one, $R$ parameter fixed to $1$, inclination angle fixed to 
$19\arcdeg$, photon index set to be equal to the slope of the cutoff power law, and energy cutoff fixed to $100\,$keV.
As expected due to the lack of
data above $10\,$keV, the errors on the reflection component normalization are rather high.
The \feka\ line parameters are: $E = 6.41 \pm 0.03\,$keV, $\sigma=26\errUD{69}{20}\,$eV, 
$F=1.50\errUD{0.64}{0.44}\times 10^{-5}\,$\normGAUSS, $\mbox{EW}=24\errUD{12}{8}\,$eV ($\chidof=1206.6/1143$).
As the reflection component is poorly constrained, we again bounded its normalization to that of the emission line.
We also added a narrow \fekb, bounded to the \feka\ as before.
Thus we obtained a best-fit parameters
consistent with the \suzaku\ one: cutoff power law and reflection continuum photon index $\Gamma =
1.58\pm 0.01$, reflection strength $\pedix{N}{pexrav}/\pedix{N}{cutoffpl}=0.19\errUD{0.05}{0.04}$, 
$\nhsym=(1.04 \pm 0.01) \times 10^{22}\,$\nh.
The \feka\ line parameters are: $E = 6.41 \pm 0.03\,$keV, $\sigma=89\errUD{53}{51}\,$eV, 
$F=2.34\errUD{0.59}{0.53}\times 10^{-5}\,$\normGAUSS, $\mbox{EW}=38\errUD{11}{9}\,$eV with respect to the observed 
continuum.
In particular, we note a lack of an evident
response of the line to the increase in the primary emission. 
Clearly, with only two observations
we cannot determine the time lag, or compare with the limit found
with the \rxte\ monitoring by \citet{chatt11}; as previously
mentioned, the correlation found between X-ray flux and \feka\ line intensity 
shows a lag less then $90$ days, a result inconsistent with a torus origin for
the bulk of the emission, although not with the BLR.
We note that, despite the different and more complex model derived here, also 
in our case we found that the EW decreases with the increasing of the flux 
(see their fig.~8).

\begin{figure}
  \centering
 
\resizebox{0.95\hsize}{!}{\includegraphics[angle=270]{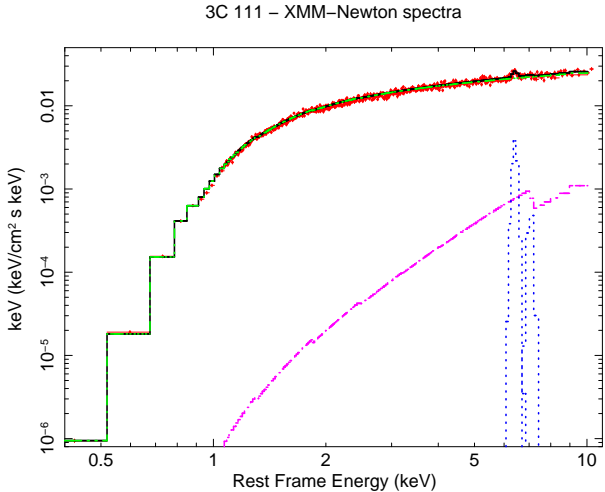}}
  \caption{\xmm\ EPIC spectra of \treca.
  The black lines represent the total best-fit model: the underlying continuum
  is modelled
  with a dominant cutoff power-law component (dash-dot-dot-dotted green line) plus a weak reflection component 
  (dash-dotted magenta line).
  The iron K line complex is composed by a \feka\ and a \fekb\ lines (blue dotted lines).
  For demonstration purposes, the data have been binned to have a significance of $20\sigma$.}
  \label{fig:xmm3c111eeuf}%
\end{figure}

This model provides an acceptable description for the  broadband X-ray spectrum of \treca\ ($\chidof=1219.7/1144$).
Observed fluxes and intrinsic luminosities are summarized in Table~\ref{tab:111fllum};
the unfolded best fit to the data is shown in Fig.~\ref{fig:xmm3c111eeuf}, while the best-fit parameters are reported
in Table~\ref{tab:111bfpar}.

No clear residuals in absorption at $\sim 7-8\,$keV are present.
Adding an absorption line with energy and width fixed to the values previously found, we obtained for the 
$E=7.23\,$keV and the $E=8.76\,$keV an upper limit to the EW of $-8.5\,$eV and $-11.1\,$eV, respectively,
formally consistent within the errors with the \suzaku\ results.

Finally, 
even though the previous model already provides a good description of the \xmm\ data,
taking into account the \suzaku\ results, we tested the \xmm\ data for the presence of possible outflowing gas.
When the {\tt XSTAR} table previously used is applied to the \xmm\ data, the gas parameters best-fitting the \suzaku\
spectra result to be inadequate to describe the EPIC data ($\chidof=1234.1/1144$): 
the lack of absorption feature at high energies, in
spectra with $S/N$ good enough to detect it, may suggests that the outflowing
gas has varied.
To further assess the possible variability of this highly ionized
absorber, we fixed its ionization and velocity to the \suzaku\ best fit
values.
Allowing only the column density to vary does not improve the fit; the parameter is unconstrained, with a $90$\% 
confidence interval between $0$ and $\pedap{N}{H}{ion}=1.2\times 10^{23}\,$\nh.
This suggests again the variable nature of this disk wind.
Allowing all the wind parameters to vary, a lower column density $\pedap{N}{H}{ion}=2.1\errUD{1.0}{0.8}\times
10^{21}\,$\nh, lower ionization parameter
$\log\xi = 2.80\errUD{0.22}{0.20}\,$erg~cm~s$^{-1}$, and zero outflow velocity $\pedix{v}{out}< 0.001\,c$, are required
($\chidof=1201.1/1141$).
The improvement in the fit with respect to the model without absorber is $\Delta\chi^2=18.6$ for $3$ degree of freedom 
less. 
This improvement is not due to single absorption
features, which are unexpected for similar values of $\log\xi$ and \pedap{N}{H}{ion},
but instead is due to a better parameterisation of the
spectral curvature due to the absorption.

%

\subsubsection{The RGS spectra}\label{sect:an3c111rgs}

In order to investigate the emission feature at $0.8-0.9\,$keV detected by \suzaku, we looked at the RGS data. 
To increase the statistics, we also reduced and analysed the RGS data obtained in March~2001; exposure times after
cleaning and count rates are reported in Table~\ref{tab:obslog}.
As anticipated, we rebinned the RGS spectra in constant wavelength bins at
twice the instrument spectral resolution ($\Delta\lambda=0.2\,$\AA) and we used the 
C-statistic \citep{cash79} available in XSPEC for the spectral fit: with this 
choice of binning, there are fewer than $20$ counts per resolution bin.

A first inspection of the RGS1\&2 data reveals that for both observations the spectra are background-dominated
below $\sim 0.65\,$keV.
We then considered in our analysis only data between $0.65$ and $2\,$keV.

%
\begin{figure}
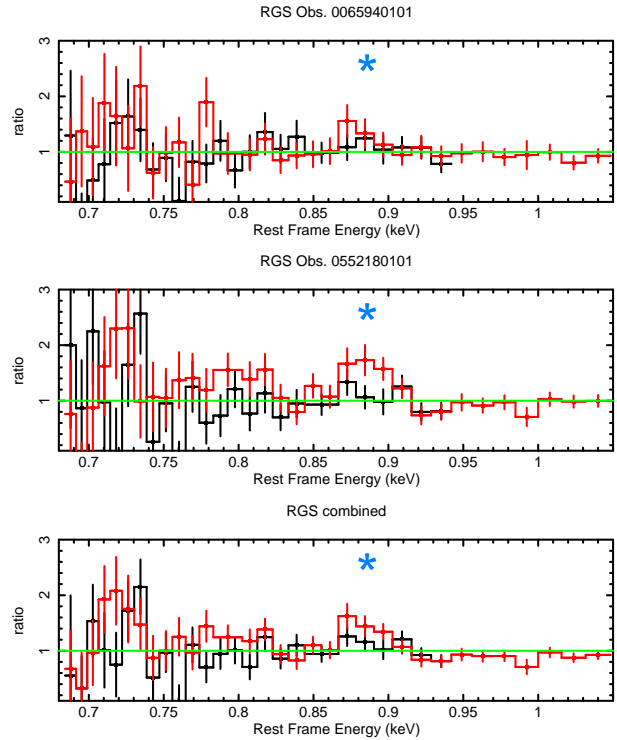

  \centering
  \resizebox{0.95\hsize}{!}{\includegraphics[angle=270]{f8a.ps}}
  \resizebox{0.95\hsize}{!}{\includegraphics[angle=270]{f8b.ps}}
  \resizebox{0.95\hsize}{!}{\includegraphics[angle=270]{f8c.ps}}
  \caption{RGS background-subtracted data-to-model ratio for an absorbed power law, in the energy range
   $0.68-1.05\,$keV.
  From top to bottom, Obs. 0065940101, Obs. 0552180101, combined spectra.
  The blue star marks the position of the emission line at $\sim 0.89\,$keV.
 }
  \label{fig:rgs3c111ra}%
\end{figure}

The two observations show a rather similar spectral shape, although with a decrease in flux of a factor of $\sim 1.5$ 
(from the oldest to the most recent).
In Fig.~\ref{fig:rgs3c111ra}, {\it upper} and {\it middle panels}, we show the data/model ratio for the two
observations when an absorbed power law is applied, with $\Gamma$
fixed to the value found for the EPIC data, $1.58$ (see Sect.~\ref{sect:an3c111xmm}).

We then combined the two observations (see Fig.~\ref{fig:rgs3c111ra}, {\it lower panel}).
Fitting the combined spectra with the same absorbed power law, we obtained a reasonable description of the RGS 
continuum
($C/\mbox{d.o.f.}=171.9/105$ for $108$ total bins).
However, linelike residuals are present between $0.85$ and $0.95\,$keV.

We then added to the AGN baseline continuum an unresolved line in emission, modelled with a Gaussian component;
the improvement in the fit due to the addition of this component is
$\Delta C/\Delta\mbox{d.o.f.}=15.2/2$.
When allowing the width to vary, the line is still unresolved.
The emission line is detected at an energy of 
$E = 0.890\pm 0.002\,$keV, with a flux of $F = 1.43\errUD{0.71}{0.61}\times
10^{-4}\,$\normGAUSS\ and an equivalent width of $\mbox{EW}=8\pm 1\,$eV.
The line parameters from RGS1 and RGS2 are consistent within the errors.

The most likely identification of this feature is the \oviii~RRC at $0.872\,$keV, although cannot be excluded a 
contribution due to the  
He-like triplet of \ion{Ne}{ix}, at energies\footnote{http://physics.nist.gov} 
$0.905\,$keV, $0.915\,$keV, and $0.922\,$keV.
The quality of the data prevents us from further investigating the properties of the line.


\section{Discussion}\label{sect:discuss}

In this paper we presented the analysis of our \suzaku\ observation of \treca, as well as archival 
unpublished \xmm\ spectra of the BLRG.

The \suzaku\ and \xmm\ emission of \treca\ is characterized by a hard
continuum, showing weak reprocessing features (iron K line and reflection
component).
The continuum is rather flat, $\Gamma\sim 1.6$, compared to the values
typically found for RQ \citep[$\Gamma\sim 1.9$; e.g.][]{reeves00,piconcelli05,mateos10} 
and BLRGs \citep[$\Gamma\sim 1.7$; e.g.][]{zdziarski01,grandi06}.
The broad-band shape observed by \suzaku\ and \xmm\ is in good agreement; the
$0.5-10\,$keV flux changes of a factor $2.5$ in the $6\,$months separating the two observations.
Such a strong flux variability is not surprising in this source. 
The power spectral density analysis of the \rxte\ monitoring between $2.4$ and $10\,$keV presented by \citet{chatt11}
reveals a break timescale of $\sim 13\,$days; flares as strong as a factor of $2-3$ in flux are observed in 
$\sim 1\,$month.
Previous \xmm\ and \rxte\ observations caught the source in a higher flux state than \suzaku\ (by a
factor of $3$ in the $2-10\,$keV energy range) as well as in the new \xmm\ data.
Again, the X-ray spectral shape does not change significantly: $\Gamma \sim 1.63-1.75$, $R \lesssim 0.3$
\citep{lewis05}.
Variations in the intensity of the emission are observed up to at least $\sim 100\,$keV, as implied 
by the comparison between the HXD-PIN emission and the BAT averaged spectrum.
Despite the continuum variation, the fluxes of the $6.4\,$keV \feka\ line, 
clearly detected in both \xmm\ and \suzaku\ 
datasets, are consistent.

\subsection{Emission and absorption features}

In addition to the \feka-\fekb\ complex, the most significant features detected in the XIS spectra are two absorption 
lines at $E\sim7.24$ and $\sim 8.77\,$keV, respectively.
The presence of such absorption features are not unique to \treca: 
an analysis of $5$ nearby, X-ray bright BLRGs
shows that these absorption lines are detected in $60$\% 
of the sources \citep[where a detailed analysis of these features is discussed]{tombesi10}.
Their likely interpretation as blue-shifted iron K lines implies an origin from 
highly ionized gas outflowing with $\pedix{v}{out}\sim 0.04\,c$, probably
related with accretion disk winds/outflows.
From the ionization parameter $\xi\equiv \pedix{L}{ion}/(n\apix{R}{2})$ and the column density of the outflowing gas,
under the reasonable assumption of thickness of the clouds $\Delta R = \nhsym/n$ lower than the distance, $\Delta R
\ll R$, we can estimate the launch radius.
Adopting the absorption-corrected ionizing
luminosity between $0.0136\,$keV and $13.6\,$keV as obtained from the \suzaku\ data ($2.5\times 10^{44}\,$\lum), we
found a sub-parsec distance from the central BH, $R < 9\times 10^{15}\,$cm.

Assuming the mean black hole mass estimated by \citet{chatt11}, $\pedix{M}{BH}=1.8\times 10^{8}\,\pedix{M}{\sun}$, the
escape velocity at this radius is a factor $2$ higher than the outflow velocity, $\pedix{v}{esc}\sim 0.077\,c$.
The highest black hole mass estimated for \treca, $\pedix{M}{BH}=3.6\times 10^{9}\,\pedix{M}{\sun}$ 
\citep[see Table~\ref{tab:src}]{marchesini04}, would imply an even worse situation, with an escape velocity a factor 
$10$ higher than \pedix{v}{out}.
The outflow is then unable to escape from the central region where is produced, and the absorber is most likely 
unstable.
Indeed, at the time of the \xmm\ observation the outflowing gas has varied.
No absorption features are detected in the EPIC spectra.
Testing a photoionized absorber as seen in the \suzaku\ data, we found a steady absorber
with lower column density and lower ionization parameter.

Finally, the XIS spectra also show an emission feature at $\sim 0.9\,$keV;
the most commonly observed features at energies around $E=0.89\pm 0.01\,$keV are the \oviii~RRC at $0.872\,$keV or the
\neix~triplet lines at 
$0.905\,$keV, $0.915\,$keV, and $0.922\,$keV;
at the XIS resolution these features are unresolved.
An emission line is also detected in the combined RGS spectra of \treca, with consistent energy, $E = 0.890\pm
0.002\,$keV.
The measured fluxes are consistent within the errors:
$1.43\errUD{0.71}{0.61}\times 10^{-4}\,$\normGAUSS\ and $2.09\errUD{0.88}{0.85}\times 10^{-4}\,$\normGAUSS\ for \xmm\
and \suzaku, respectively.
Similar features, produced by photoionized gas, have been detected in 
two other FRII radio galaxies, the BLRG 3C~445
\citep[e.g.][]{grandi07,sambruna07} and the NLRG 3C~234 \citep{piconcelli08}.

\subsection{From X-ray to $\gamma$-ray}

The comparison presented here between HXD-PIN emission and the BAT averaged spectrum at high energies, and \xmm\ and 
XIS data at lower energies, confirms the strong variability of \treca, which could be tentatively ascribed to the jet
variability.
Only broad-band monitoring observations will allow us to confirm or rule out this possibility.
No clear roll-over is detected in the \suzaku\ data, which can be explained by the presence of such a jet component.
Indeed, if the GSO detection is robust, there is a marginal evidence of the emerging of a jet component 
at energies higher than $10\,$keV,
although the present data do not allow us to firmly establish its relative contribution with respect to the primary
continuum.
The strong variability observed in the X-ray flux
together with the flat photon index of the continuum, 
could suggest dilution of the disk emission due to a variable jet contribution. 
In principle, data with higher spatial resolution could allow us to investigate the jet contribution below
$10\,$keV, resolving some knots that are blended for \xmm\ (and even more for \suzaku); indeed, X-ray jet emission may
arise on a range of scales, as the one-sided radio jet is detected at mas to arc-minutes.
However, images from a \chandra\ observation \citep{hogan11}
show that all but one the visible knots along the jet fall
outside the extraction region of the \xmm\ spectra, and none of them is bright enough to be detected in the automatic
source detection and analysis performed with {\tt XASSIST}\footnote{http://xassist.pha.jhu.edu/zope/xassist}.
The lack of unresolved bright knots in the \xmm\ spectra, that correspond to a higher flux level, make unlikely that the
variation is jet-related.

%
\begin{table}
\begin{minipage}[t]{\columnwidth}
\caption{{\bf Spectral indices and fluxes from the combined fit to the X-ray SED.}}
\label{tab:111sed}
\begin{center}
{\footnotesize
\begin{tabular}{r@{\extracolsep{0.15cm}} l@{\extracolsep{0.2cm}} c@{\extracolsep{0.2cm}} c@{\extracolsep{0.2cm}} c}
 \hline \hline
   Satellite & (Year) & $\Gamma$ & \multicolumn{2}{c}{Flux$^{a}$ [$10^{-11}\,$\flux]} \\
   \cline{4-5}
   &  &  & $0.5-10\,$keV & $10-200\,$keV\\
 \hline
   \sax\      & (1998)     &  $1.67\pm 0.06$           & $2.84$ & $7.42 $ \\  
   \xmm\      & (2001)     &  $1.68\pm 0.01$           & $7.43$ & $15.78$ \\  
   \swift\    & (58-month) &  $1.57\pm 0.09$           & $3.90$ & $11.41$ \\  
   \suzaku\   & (2008)     &  $1.61\pm 0.01$           & $2.22$ & $6.89 $ \\  
   \integral\ & (2008)     &  $0.95\errUD{0.43}{0.40}$ & $1.08$ & $12.35$ \\  
   \xmm\      & (2009)     &  $1.61\pm 0.01$           & $5.45$ & $14.05$ \\  
 \hline
\end{tabular}
}
\end{center}       
{\footnotesize  Errors are quoted at the 90\% confidence level for 1 parameter 
of interest ($\Delta\apix{\chi}{2}=2.71$). 
$^{a}\,$Observed fluxes.
} \\
\end{minipage}
\end{table}
%

%
\begin{figure*}
  \centering
\vspace{0.5cm}
  \resizebox{0.9\hsize}{!}{\includegraphics[angle=90]{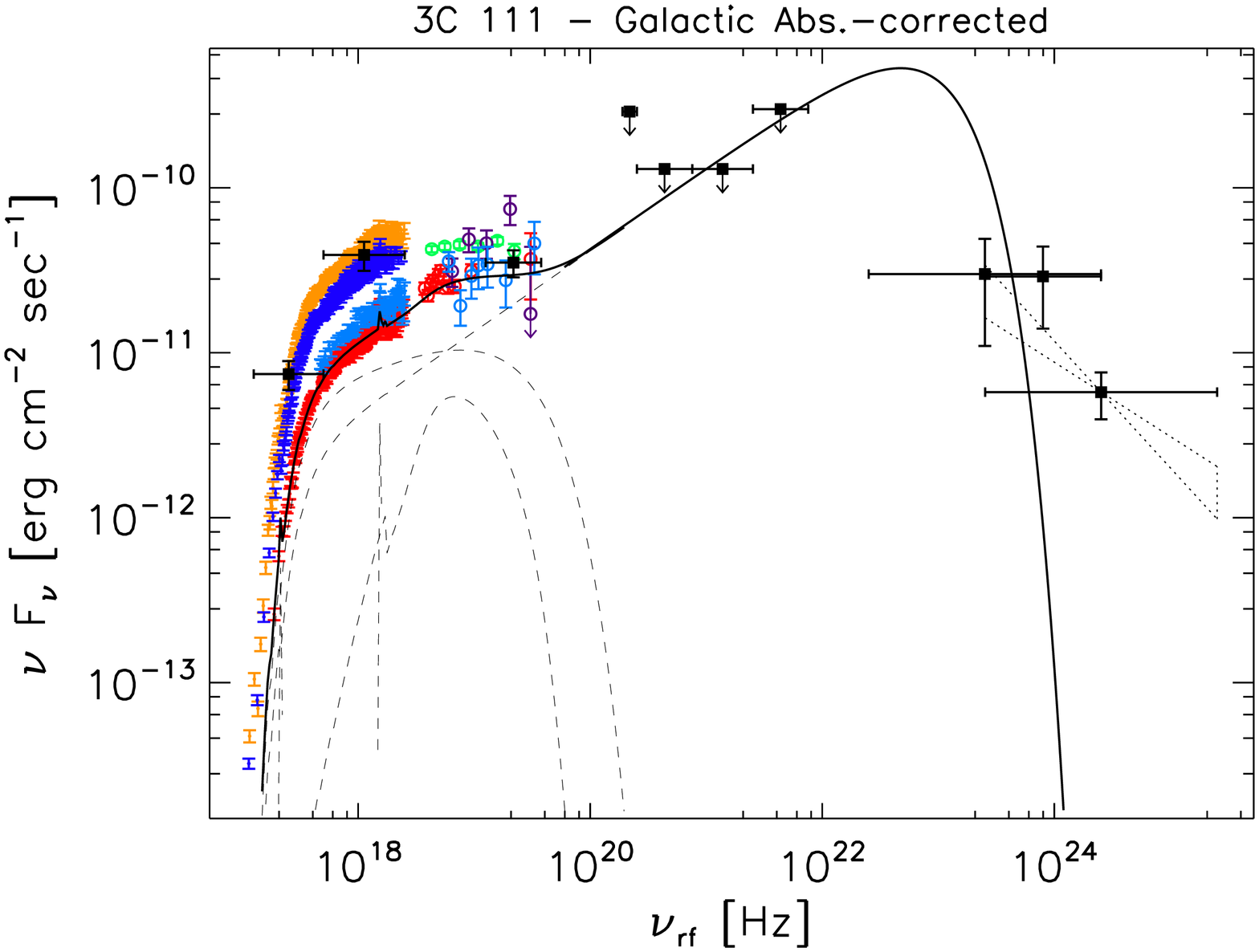}}
  \caption{Rest-frame high-energy SED of \treca.
  Red symbols are \suzaku\ spectra (XIS03, PIN, and GSO); above $10^{18}\,$Hz, green and purple 
  circles are BAT and INTEGRAL data, respectively. \sax\ MECS and PDS spectra are plotted in light blue; orange symbols
  are \xmm\ MOS data from 2001, while blue spectrum is the \xmm\ MOS2 spectrum analysed
  in this paper.
  Black filled squares are taken from literature \citep{hartman08,abdo10}: from low to high energies, \asca, {\it CGRO} 
  (\osse, \comptel, and \egret\ instruments), and \fermi\ data.
  Data are corrected for local absorption.
  Dashed black lines are the components of the best-fit model for the \suzaku\ data when the jet component is included;
  the solid line corresponds to the total model, with a cut off at $\sim 10\,$MeV applied.}
  \label{fig:3c111sed}%
\end{figure*}

In the $\gamma$-ray band, an \egret\ identification of \treca\ as a $\gamma$-ray source
has been suggested, although with some debate on the
real contribution of \treca\ to the total $\gamma$-ray emission of the \egret\ 
counterpart \citep{sguera05,hartman08}.
Recently, an association with a \fermi\ source has been reported by 
\citet{abdo10}.
The fluxes detected in the $\gamma$-ray band are:
$\pedix{F}{0.75-3\,\mbox{\scriptsize MeV}}<4.2\times10^{-10}\,$\flux\ and 
$\pedix{F}{3-30\,\mbox{\scriptsize MeV}}<4.3\times10^{-10}\,$\flux\ (\comptel), and
$\pedix{F}{0.1-10\,\mbox{\scriptsize GeV}}\sim3\times10^{-11}\,$\flux\ 
[\egret, both on-board the \cgro\ ({\it CGRO}); \citealt{hartman08}]; 
$\pedix{F}{1-100\,\mbox{\scriptsize GeV}}\sim6\times10^{-12}\,$\flux\ \citep[\fermi;][]{abdo10}.
These fluxes are highly above the value predicted by the best-fit model to the $0.6-70\,$keV \suzaku\ data without a 
jet component.
This supports the hypothesis of high-energy emission due to the jet in \treca.
Recently, \citet{kataoka11} ascribed the \fermi\ detection to a pc-scale
relativistic emission from the jet.
Extrapolating  the dual power law model, including a possible jet component (see Sect.~\ref{sect:an3c111suz}), to high 
energies, we found for the jet component $\pedap{F}{3-30\,\mbox{\scriptsize MeV}}{jet}=4.9\times10^{-10}\,$\flux.
Taking into account the typical variability of jet emission, this flux is in substantial agreement with the upper limits
obtained in the \comptel\ energy range.
Qualitatively, a peak at $\sim 100\,$MeV would cause the observed jet component to be compatible with the observed 
emission above $0.1\,$GeV.

This is evident looking at Fig.~\ref{fig:3c111sed}, where we present the high-energy historical SED
for \treca\ from X-ray to $\gamma-$ray energies.
We used the \suzaku, \xmm, and \swift\
observations analysed in this paper together with \asca, \xmm\ MOS (from 2001), \sax\ (from 1998), \integral\ 
(from 2008), {\it CGRO}, and \fermi\ data.
For safe of clarity, in Fig.~\ref{fig:3c111sed} we do not show XIS1 and pn (from 2009) data.
We note that the \swift\ spectrum is averaged over more than $4\,$years.
The strong variability previously mentioned is clearly evident when all the datasets are plotted together.
In order to understand the driver of the variability, we fitted all the datasets together,
adopting the best-fit continuum model of the \suzaku\ data, \ie\ an absorbed cutoff power law plus a 
reflection  component and the iron K complex; since the comparison between \suzaku\ and \xmm\ spectra shows that the 
lines do not vary with the continuum, we fixed the reprocessing features to the best fit value found for \suzaku.

As a first step, we allowed only the relative normalization of the primary power law to vary between the observations.
For observations extending down to a few keV, also the column density was allowed to vary.
This fit already provides a 
reasonable
representation of the SED ($\chidof=3177.6/2429$), 
our aim being to investigate the variations in the main spectral properties (continuum slope and emission
intensity), and not to find a detailed description of all the spectra.
However, to further 
investigate the observed variability, we also allowed the power-law photon index to vary; the observed change of 
intensity and spectral shape is summarized in Table~\ref{tab:111sed}
($\chidof=2972.8/2423$).
We obtained rather similar values of the photon index, spanning a range 
from $\Gamma=1.57\pm 0.09$ (\swift) to $\Gamma=1.68\pm 0.01$ (\xmm\ from 2001).
The only outlier is the \integral\ observation, for which we obtain $\Gamma=0.95\errUD{0.43}{0.40}$.
This is probably due to the low quality of the data, confirmed by the poor constraints to the value of $\Gamma$, coupled
with the high-energy cutoff applied, $\pedix{E}{c}=100\,$keV.
Fitting the data with a simple power law would increase
a little the best-fit value of the photon index, although with strong residuals at the high-energy end of the spectrum.
The column density varies between $\sim 1.08\times10^{22}\,$\nh\ (\xmm\ from 2009) and $\sim 1.50\times10^{22}\,$\nh\
(\sax\ from 1998).
As noted before, we observe significant variability in the $0.5-10\,$keV flux, from $\sim 2.8\times10^{-11}\,$\flux\ 
(our XIS data) to $\sim 7.4\times10^{-11}\,$\flux\ (\xmm\ from 2001).
At higher energies, the stronger variation is observed between the \suzaku\ and the \xmm\ (from 2001) data.

This analysis shows that a brightening of the source does not imply a flatter power law, thus suggesting that the main 
driver of the increase in the flux may not be a strong increase in the jet component.
Although our analysis is limited by the number of spectra, we note that the photon indices are well constrained thanks 
to the use of soft X-ray data as well as data above $10\,$keV.
After excluding the \integral\ point, the change in slope with the flux found here, adopting a more complex model, is 
marginally consistent with that obtained from the \rxte\ monitoring \citep[see their fig.~8]{chatt11}, 
which shows a similar trend albeit with a larger $\Delta\Gamma$ and covering a large range in the $2-10\,$keV flux.

In our observational program, we have BLRGs showing a spread of the jet angle with respect to the line of sight.
Despite the strong uncertainties in the orientation, the closest candidate to be compared to \treca\ is \trecb.
Recently, from high-quality \suzaku\ data \citet{sambruna11} estimated for this BLRG a disk inclination of
$\sim 25\arcdeg-30\arcdeg$; at bigger scales, radio observations found a lower limit to the inclination angle of $\sim
15\arcdeg$ \citep{eracleous98}.
The best estimate of the jet inclination for \treca\ is $\sim 19\arcdeg$, with a proposed range of
$10\arcdeg<i<26\arcdeg$ \citep{kadler08,lewis05}.
Despite this similarity, the X-ray emission of the two sources is remarkably different, with \trecb\ showing an X-ray
spectrum more similar to that observed from Seyfert galaxies.
This implies that the relative importance of the jet is not related only to geometrical properties (\ie, orientation
with respect to the line of sight).
Different velocities, or a bent jet, can play an important role.

Finally, the X-ray emission of \treca\ appears significantly different 
from that of a typical Seyfert galaxy in many aspects, \eg\ flatter intrinsic emission and weak reflection features.
On the other hand, its time variability shows properties that correlate with the black hole mass, following the same
scaling law observed in stellar-mass black hole X-ray binaries and Seyfert galaxies, suggesting that a similar accretion
process is powering these different systems \citep{chatt11}.
It is worth noting the large dispersion of intrinsic X-ray
properties shown by RL AGN \citep[and references therein]{sambruna09}, similar to the one found out for the RQ sources.
The overlapping of the distributions of photon index and reflection features 
observed in both 
classes supports the idea of common accretion structure, with a second 
parameter in addition to
the accretion rate that determines the jet production efficiency.
The most likely candidate is the black hole spin, although how does it works in 
producing the
observed RLs-RQs dichotomy, or the FRIs-FRIIs division, is still strongly 
debated 
(e.g., the ``spin paradigm'' coupled with intermittent jet activity, 
\citealt{sikora07}; or the
``gap paradigm'', \citealt{garofalo09,garofalo10}).
In particular, in the heuristic scenario proposed by Garofalo et al., collimated jets coupled with intermediate 
accretion efficiency, as observed in \treca\ (see Table~\ref{tab:src}), can occur for retrograde systems.
This configuration also implies a larger size of the gap region between the inner edge of accretion disks and the black
hole horizon, size that is connected to the energetics of the disk itself.
The weakness of reflection features is a natural consequence of such a configuration; moreover, larger gap regions can 
limit the presence of disk winds \citep{kuncic04,kuncic07}.
Unstable outflows, weaker than in RQ AGN, could be therefore compatible with High Excitation FRII states, such in 
\treca\ \citep{buttiglione10}, while at the extreme end, radiatively inefficient
accretion flows in Low Excitation Radio Galaxies ($L/\pedix{L}{Edd}\sim10^{-(5-7)}$) might inhibit the launch of disk 
winds.


\section{Summary}\label{sect:summ}

In this paper we presented the X-ray spectra of \treca, 
obtained with \suzaku\ as part of our observational program devoted to study 
the high-energy emission of the brighter BLRGs.
A recent \xmm\ observation of $\sim 120\,$ksec is analysed as well.

The \treca\ emission extends up to $\sim200\,$keV, with a rather flat 
continuum; 
in spite of this, no signatures of jet emission are visible below $10\,$keV.
\suzaku\ observed the source at a minimum flux level, as shown by the \rxte\ monitoring.
In the $6\,$months separating
the \suzaku\ and \xmm\ observations, the flux increases by a factor of $2.5$, a level of variability which is not
unexpected in this source.
We confirm the weakness of the reflection features, as found in previous observations.
An iron K complex is clearly detected in both datasets, with a rather low EW.
The intensities of the line detected by \xmm\ and \suzaku\ are consistent, with a lack
of immediate response of the line properties to the continuum variation.
In addition to the iron K complex, absorption features due to a highly ionized, ultra-fast, nuclear outflowing gas 
are detected in the XIS data.
At the time of the \xmm\ observation, the gas has varied: adopting the photoionized absorber as seen in the \suzaku\
data, then a lower column density and lower ionization state are required.
Both XIS data and RGS spectra show the presence of a narrow line at $\sim 0.89\,$keV, 
which is most likely identified with either the \oviii~RRC or the \neix~triplet lines.

The lack of detection of a roll-over in the primary emission is probably due to the appearance of the
jet as a dominant component in the hard X-ray band, as suggested by the detection with the GSO on-board \suzaku\ above 
$\sim 100\,$keV.
From a qualitative point of view, the emission observed by \egret\ and \fermi\ above $0.1\,$GeV is compatible with such
a jet component assuming that this peaks at energies of $\sim 100\,$MeV.
If the detection is confirmed, the emergence of a jet component can explain differences observed in the high-energy
emission of BLRGs similar in other aspects.
The relative importance of the jet component is not simply related to the system inclination, but can be associated to
the velocity of the jet; a change of the opening angle of the
jet passing from the nuclear region, where the observed  X-ray emission is
produced, to the most distant regions, where the radio emission is observed, may also
play a role.

Finally, we presented the high-energy historical SED for \treca, from X-ray to $\gamma$-ray. 
Our qualitative analysis suggests that the strong variability observed 
for \treca\ is probably driven by a change in the primary continuum.
This implies that simultaneous broad-band X-ray and $\gamma$-ray
monitoring is needed to unambiguously 
constrain the parameters of the disk-jet system,
and correctly understand the contribution of the 
different components (\ie, disk emission, reflection, jet) to the total 
emission.


\section*{Acknowledgments}

This research has made use of data obtained from the High Energy
Astrophysics Science Archive Research Center (HEASARC), provided by
NASA's Goddard Space Flight Center, and of the NASA/IPAC Extragalactic
Database (NED) which is operated by the Jet Propulsion Laboratory,
California Institute of Technology, under contract with the National
Aeronautics and Space Administration. 
Based on observations obtained from the \suzaku\ satellite, 
a collaborative mission between the space
agencies of Japan (JAXA) and the USA (NASA), and
with \xmm\ (an ESA science mission with instruments and 
contributions directly funded by ESA Member States and the USA, NASA).
We warmly thank the referee for her/his suggestions that significantly improved the paper.
We are grateful to M. Ceballos, R. Saxton and S. Sembay for their help in handling 
the \xmm\ data problems.
We warmly thank V. Bianchin for reducing the \integral\ data.
R.M.S. acknowledges support from NASA through the \suzaku\ program. 
V.B. acknowledges support from the UK STFC research council.
L.B. acknowledges support from the Spanish Ministry of Science and Innovation 
through a ``Juan de la Cierva'' fellowship.  
Financial support for this work was provided by the Spanish Ministry of 
Science and Innovation, through research grant AYA2009-08059.

\bibliographystyle{mn2e} 

\label{lastpage}


\end{document}